\documentclass[12pt]{article}
\usepackage{multicol}
\usepackage{graphicx}
\usepackage{amsmath}
\usepackage[a4paper]{geometry}
\usepackage{adjustbox}
\usepackage{hyperref}
\usepackage{rotating}
\usepackage{xcolor}

\def \be  {\begin{equation}}
\def \ee  {\end{equation}}
\def \ee  {\end{equation}}
\def \bea {\begin{eqnarray}}
\def \eea {\end{eqnarray}}

\usepackage{longtable}
\usepackage{caption}
\usepackage{longtable}
\usepackage{array}
\usepackage[aboveskip=2pt,belowskip=4pt]{caption}

\newcolumntype{L}[1]{>{\raggedright\arraybackslash}p{#1}}

\begin{document}

\begin{center}

{\Large Exploring Thermalization and Multi-Freeze-Out Effects in Pb–Pb Collisions Based on Tsallis $p_T$ Distributions} \\

\vspace{10mm}
Haifa I. Alrebdi$^{1,}${\footnote{Corresponding author: hialrebdi@pnu.edu.sa}},
Muhammad Ajaz$^{2}${\footnote{Corresponding author: ajaz@awkum.edu.pk}},
Murad Badshah$^{2}$, 
Mohammad Ayaz Ahmad$^{3}$ 

{\small\it
$^1$ Department of Physics, College of Science, Princess Nourah bint Abdulrahman University, P.O. Box 84428, Riyadh 11671, Saudi Arabia\\
$^2$ Department of Physics, Abdul Wali Khan University Mardan, Mardan 23200, Pakistan \\
$^3$ Department of Mathematics, Physics and Statistics, Faculty of Natural Sciences, University of Guyana, Georgetown 101110, Guyana, South America\\
}

\end{center}

\begin{abstract}

This study investigates the transverse momentum ($p_T$) distributions of $\pi^{\mp}$, $K^{\mp}$, $p$($\bar{p}$), $K_s^0$, and $\Lambda$ in various centrality classes of Pb–Pb collisions at $\sqrt{s_{NN}} = 2.76$ TeV. The experimental spectra are analyzed using the Tsallis non-extensive distribution, from which the effective temperature ($T$), non-extensive parameter ($q$), and mean transverse momentum ($\langle p_T \rangle$) are extracted for each particle species and centrality bin. To disentangle thermal and collective effects, the mean kinetic freeze-out temperature ($\langle T_0 \rangle$) is obtained from the intercept of the $T$ versus mass relation, while the average transverse flow velocity ($\langle \beta_T \rangle$) is extracted from the slope of $\langle p_T \rangle$ versus mean moving mass for pions, kaons, and protons.
The results show that $T$ increases and $q$ decreases with centrality, indicating a hotter and more equilibrated system in central collisions. A clear mass dependence of $T$ supports the presence of a multi-freeze-out scenario, with heavier particles decoupling earlier. Both $\langle T_0 \rangle$ and $\langle \beta_T \rangle$ rise from peripheral to mid-central collisions before saturating toward central events, which may suggest the onset of collective behavior or changes in freeze-out dynamics. These observations provide new insights into the thermal and dynamical properties of the medium created in heavy-ion collisions at the LHC.
\end{abstract}



\section{Introduction}
Heavy-ion collisions provide a unique opportunity to study matter under extreme conditions and to create the quark–gluon plasma QGP, a deconfined state of quarks and gluons. Indications of QGP formation were observed in earlier experiments at the CERN SPS \cite{sps}, and compelling evidence came from the RHIC experiments \cite{ARSENE20051, ADAMS2005102}, which reported phenomena consistent with a new state of partonic matter. With the advent of the LHC, these collisions enter a new energy regime aimed at a more precise characterization of QGP properties.

In ultra-relativistic heavy-ion collisions, the created matter exhibits strong collective behavior, flowing with an extremely low viscosity to entropy ratio like a nearly perfect liquid \cite{collectivity}. This collectivity can be studied via the transverse momentum ($p_T$) distributions of the produced particles \cite{hia, Aj}. The shapes of identified-particle $p_T$ spectra reflect the system’s bulk properties, such as the temperature and flow velocity at kinetic freeze-out, and thus provide crucial information for interpreting QGP-related observables. Any signal originating from the QGP phase must be considered in conjunction with the system’s space-time evolution through expansion and cooling.

Relativistic hydrodynamic models have been very successful in reproducing the low-$p_T$ spectra and flow patterns observed in heavy-ion collisions \cite{hydroModels}, suggesting that the bulk of the system reaches local thermal equilibrium. The transverse momentum distributions of hadrons contain information about the transverse expansion velocity of the fireball and the temperature at the moment of thermal decoupling freeze-out. As the hot system expands and cools, it undergoes a transition from a partonic QGP phase back to a hadronic phase. Inelastic collisions cease at the chemical freeze-out point, fixing the relative hadron abundances. Elastic collisions continue to build collective flow until kinetic freeze-out, when even elastic interactions cease and the final momentum distributions are set. It is generally assumed that the late hadronic stage has only a minor effect on particle yields no significant change in abundances \cite{PhysRevLett.86.2980}. Notably, the chemical freeze-out temperature $T_{ch}$ extracted at the LHC approximately 155–160~MeV is found to be close to the predicted QCD phase transition temperature \cite{ANDRONIC2009142, BRAUNMUNZINGER200461}, suggesting that chemical freeze-out occurs near the quark–hadron phase boundary. Thermal model fits of hadron yields have indeed been successful over a broad range of collision energies, yielding a common $T_{ch}$ that rises with beam energy and saturates around 160~MeV for top RHIC and LHC collisions \cite{ANDRONIC2009142, Andronic2011}. However, the LHC data have also revealed tensions with this equilibrium picture. In particular, ALICE observed that the antiproton yield in central Pb–Pb collisions is significantly lower than expected for a single freeze-out temperature of $\sim$160~MeV \cite{PhysRevLett.109.252301}. The measured proton-to-pion ratio is only about two-thirds of the thermal model prediction roughly 1.5 times smaller than expected \cite{PhysRevLett.109.252301}, indicating that some post-chemical freeze-out reprocessing such as baryon–antibaryon annihilation in the hadronic phase may be reducing the final proton abundance. Eventually, the system’s expansion leads to kinetic freeze-out at a lower temperature $T_{kin}$, when elastic collisions cease. Blast-wave fits to heavy-ion spectra indicate $T_{kin}\approx 100$~MeV for central Pb–Pb collisions at LHC energies, substantially below $T_{ch}$ \cite{acharya2020production}. This difference between $T_{ch}$ and $T_{kin}$ underlines that after chemical freeze-out the fireball continues to cool and expand, and hadrons can gain additional transverse momentum from radial flow without changing their relative abundances.

Another important observation is that in the intermediate-$p_T$ region (approx. 2–6~GeV/$c$), baryon-to-meson ratios (e.g. $p/\pi$ and $\Lambda/K^0_S$) are strongly enhanced in central Pb–Pb collisions, reaching values around unity at $p_T\sim3$~GeV/$c$ \cite{Fries2003reco, ALICELambdaKs}. This is in stark contrast to the much lower ratios seen in $pp$ collisions, and it suggests an alternate hadronization mechanism at work. Specifically, quark recombination coalescence models predict that quarks from the deconfined medium can recombine into baryons, leading to a relative enhancement of baryon production at intermediate $p_T$ \cite{Fries2003reco}. Such a mechanism, which is difficult to reproduce with purely thermal or fragmentation models, provides a natural explanation for the baryon–meson anomaly and necessitates a more differential approach to modeling the spectra.

To account for both equilibrium and non-equilibrium features of the $p_T$ spectra, a useful empirical approach is to employ the Tsallis distribution, a statistical distribution that generalizes the Boltzmann–Gibbs exponential to include deviations from thermal equilibrium. The Tsallis distribution introduces a non-extensivity parameter $q$. In the limit $q \to 1$ it reduces to the standard exponential Maxwell–Boltzmann distribution, while $q>1$ produces a harder, power-law tail. This form has been shown to describe measured $p_T$ spectra in a wide variety of collision systems from RHIC to the LHC \cite{Cleymans2012, CLEYMANS2013351, buzatu}. Physically, the parameter $q$ quantifies the degree of deviation from ideal thermal equilibrium caused by effects such as long-range correlations or intrinsic fluctuations; central heavy-ion collisions typically yield $q$ values very close to unity indicating near-equilibrium conditions, whereas smaller or more peripheral collisions exhibit larger $q$, reflecting more pronounced non-equilibrium dynamics \cite{yang2021dependence}.

The data analyzed in this work were recorded by the ALICE detector, which provides excellent tracking and particle identification PID capabilities at midrapidity \cite{ALICEdetector}. In particular, the Inner Tracking System ITS, Time Projection Chamber TPC, and Time-of-Flight TOF detectors allow identification of pions, kaons, and protons over a wide momentum range, while strange hadrons ($K^0_S$, $\Lambda$) are reconstructed via their characteristic V$^0$ decay topology. The collision centrality is determined using the VZERO scintillator detectors and a Glauber model fit to their amplitude distribution \cite{ALICEcentrality}. 

In this study, we utilize $p_T$ spectra of $\pi^{\pm}$, $K^{\pm}$, $p$ and $\bar{p}$ in Pb–Pb collisions at $\sqrt{s_{NN}}=2.76$~TeV from ALICE \cite{source1}, along with corresponding spectra for $K^0_S$ and $\Lambda$ hyperons \cite{source2}, across multiple centrality classes. We fit these spectra with the Tsallis distribution see Eq. 2 to extract the effective temperature $T$ and non-extensivity parameter $q$ for each particle species and centrality bin. From the mass-dependence of the fitted $T$ parameters, we estimate a common kinetic freeze-out temperature $\langle T_0\rangle$ for the system, and from the mean transverse momenta $\langle p_T\rangle$ of the particles we infer the average transverse flow velocity $\langle \beta_T\rangle$. Examining how these parameters vary with centrality allows us to characterize the degree of thermalization and collective expansion in different collision environments. In particular, we investigate whether $\langle T_0\rangle$ and $\langle \beta_T\rangle$ tend to saturate in the most central collisions. This behavior could signal the onset of a new dynamical regime or the approach to the limits of hydrodynamic expansion.

\section{The Experimental Data and Formalism}

The present analysis utilizes the $p_T$ spectra of identified hadrons (pions, kaons, and anti-protons) measured by the ALICE Collaboration in heavy-ion collisions at the LHC. In particular, we consider the experimental data for Pb+Pb collisions at $\sqrt{s_{\mathrm{NN}}} = 2.76$ TeV across multiple centrality classes, as reported by ALICE \cite{source1, source2}. These mid-rapidity spectra ($|y| \approx 0$ within the ALICE central barrel acceptance) have been corrected for detector acceptance, efficiency, and feed-down contributions from weak decays. The analysis covers ten centrality intervals from central 0–5\% to peripheral 80–90\%, with each interval characterized by the average number of participant nucleons $N_{\text{part}}$ and the charged-particle multiplicity density $\langle dN_{\text{ch}}/d\eta \rangle$ as determined from a Glauber model simulation. 
The statistical and systematic uncertainties of the data points are combined in quadrature on the spectra plots.

To describe the spectral shapes, we employ the non-extensive Tsallis statistical framework, which has been widely used to model particle production in high-energy collisions see, e.g., Refs.~\cite{Tsallis:1988, Tang:2009, Cleymans2012}. The Tsallis distribution is a generalization of the Boltzmann–Gibbs exponential law with an additional parameter $q$ that quantifies deviations from equilibrium. At mid-rapidity ($y \approx 0$) and zero chemical potential, one convenient form of the Tsallis distribution can be written as: 

\begin{equation}\label{eqTsallisSimple}
\frac{1}{2\pi N_{\rm ev}\,p_T}\frac{d^2N}{dp_T\,dy} = C\bigl[1 + (q-1)\tfrac{m_T}{T}\bigr]^{-1/(q-1)},
\end{equation}

where $N_{\rm ev}$ is the number of events, $m_T = \sqrt{p_T^2 + m_0^2}$ is the transverse mass (with $m_0$ the particle rest mass), $T$ is an effective temperature parameter, and $C$ is a normalization constant. The exponent $q$ is the non-extensivity parameter: in the limit $q \to 1$. Eq.~\ref{eqTsallisSimple} smoothly reduces to the standard exponential Boltzmann–Gibbs distribution, as expected for a system in thermal equilibrium. Values of $q>1$ reflect departures from equilibrium, often attributed to long-range interactions or intrinsic fluctuations in the system. The magnitude of $q-1$ can be interpreted as a measure of the degree of non-equilibrium or “temperature fluctuation” in the source. In high-energy collision data, Tsallis fits with $q>1$ have been shown to describe the power-law tails of $p_T$ spectra effectively over a broad momentum range. 

It should be noted that several equivalent formulations of the Tsallis distribution exist in the literature. In particular, a thermodynamically consistent form has been derived to ensure that pressure, energy density, and entropy are all well-defined. Its mid-rapidity form (for $\mu \approx 0$) can be written as:

\begin{equation}\label{eqTsallisConsistent}
\frac{1}{2\pi N_{\rm ev}\,p_T}\frac{d^2N}{dp_T\,dy} = C_q \, m_T \,\bigl[1 + (q-1)\tfrac{m_T}{T}\bigr]^{-q/(q-1)}.
\end{equation}

This consistent formulation \cite{Cleymans2012, CLEYMANS2013351, Buzatu:2019} satisfies standard thermodynamic relations and typically yields a somewhat lower extracted $T$ than Eq.~\ref{eqTsallisSimple}. For our purposes, we adopt Eq.~\ref{eqTsallisConsistent} to extract meaningful thermodynamic parameters. 

Collective transverse flow in heavy-ion collisions serves to harden the $p_T$ spectra. A direct Tsallis fit without flow thus yields an “effective temperature” $T$ that conflates thermal motion and radial flow. To disentangle these, one can explicitly include a flow velocity in the distribution. Following Olimov et al.~\cite{Olimov:2022MPLA, Olimov:2022Universe}, one applies a Lorentz boost to the transverse mass: replace $m_T$ by $\langle \gamma_T \rangle \bigl(m_T - \langle \beta_T \rangle\, p_T\bigr)$ with $\gamma_T = 1/\sqrt{1 - \beta_T^2}$. The resulting Tsallis-with-flow reads

\begin{equation}\label{eqTsallisConsFlow}
\frac{1}{2\pi N_{\rm ev}\,p_T}\frac{d^2N}{dp_T\,dy}
= C_{q}\,\langle \gamma_T \rangle\,\bigl(m_T - \langle \beta_T \rangle\, p_T\bigr)\,
\bigl[1 + (q-1)\tfrac{\langle \gamma_T \rangle\,\bigl(m_T - \langle \beta_T \rangle\, p_T\bigr)}{T_0}\bigr]^{-q/(q-1)},
\end{equation}

where $T_0$ is interpreted as the kinetic freeze-out temperature and $\beta_T$ the average transverse flow velocity. In principle, simultaneous fits of all hadron species using Eq.~\ref{eqTsallisConsFlow} yield a unique pair $(T_0,\beta_T)$ characterizing the system.

As an alternative, and the method we actually use here, we first fit each species individually with the \emph{flow-absent} consistent Tsallis, Eq.~\ref{eqTsallisConsistent}, to extract $T(m_0)$. Then, to extract the kinetic freeze-out temperature $T_0$ and the average transverse flow velocity $\langle\beta_T\rangle$, we perform two independent linear fits:

\begin{align}
T(m_0) &= T_0 + a\, m_0, \tag{4a} \label{eq:4a}\\
\langle p_T \rangle(\overline{m}) &= b +\overline{m}\, \langle \beta_T \rangle. \tag{4b} \label{eq:4b}
\end{align}

Here, $m_0$ is the particle rest mass, $\overline{m}$ denotes the mean moving mass for the species, and $a$ and $b$ are fit constants. 
The intercept of \eqref{eq:4a} at $m_0=0$ gives $T_0$, while the slope of \eqref{eq:4b} yields $\langle \beta_T \rangle$.
 Using two separate fits avoids possible correlations between parameters in a combined model and provides a cross-check of the extracted values.

Regardless of the specific form of the particle momentum distribution, the transverse momentum-dependent probability density function can be expressed as
 \begin{align}
f(p_T) = \frac{1}{N}\frac{dN}{dp_T}.
\end{align}
which is naturally normalized to unity,
\begin{align}
\int_{0}^{\infty}f(p_T)\ dp_T = 1.
\end{align}
The mean transverse momentum, $\langle p_T\rangle$, can then be determined directly from the fit function by employing the probability density function as
 \begin{align}
\langle p_T \rangle = \frac{\int_{0}^{\infty}p_T\ f(p_T)\ dp_T}{\int_{0}^{\infty}f(p_T)\ dp_T}.
\end{align}
Making use of Eq. (5) we get,
\begin{align}
\langle p_T\rangle = \int_{0}^{\infty}p_T\ f(p_T)\ dp_T.
\end{align}

\section{Results and discussion}
\label{sec:RsltDsc}
\subsection{Tsallis Fit Quality for Identified Hadron Spectra}
Figures ~1 and 2 demonstrate that the Tsallis distribution provides an excellent description of the transverse momentum spectra for all studied hadrons across a broad range of centralities. In Fig. ~1a--f, the measured $p_T$ spectra of $\pi^+$, $\pi^-$, $K^+$, $K^-$, $p$, and $\bar{p}$ in Pb–Pb collisions at $\sqrt{s_{NN}}=2.76$~TeV are plotted alongside our Tsallis fit curves. Similarly, Fig.~~2a and 2b show the spectra of the strange hadrons $K^0_S$ and $\Lambda$ with their respective fits. 

\begin{figure}[p!]
\vspace{-1.0cm}
\begin{center}
\vspace{-0.4cm}
\includegraphics[width=0.48\textwidth]{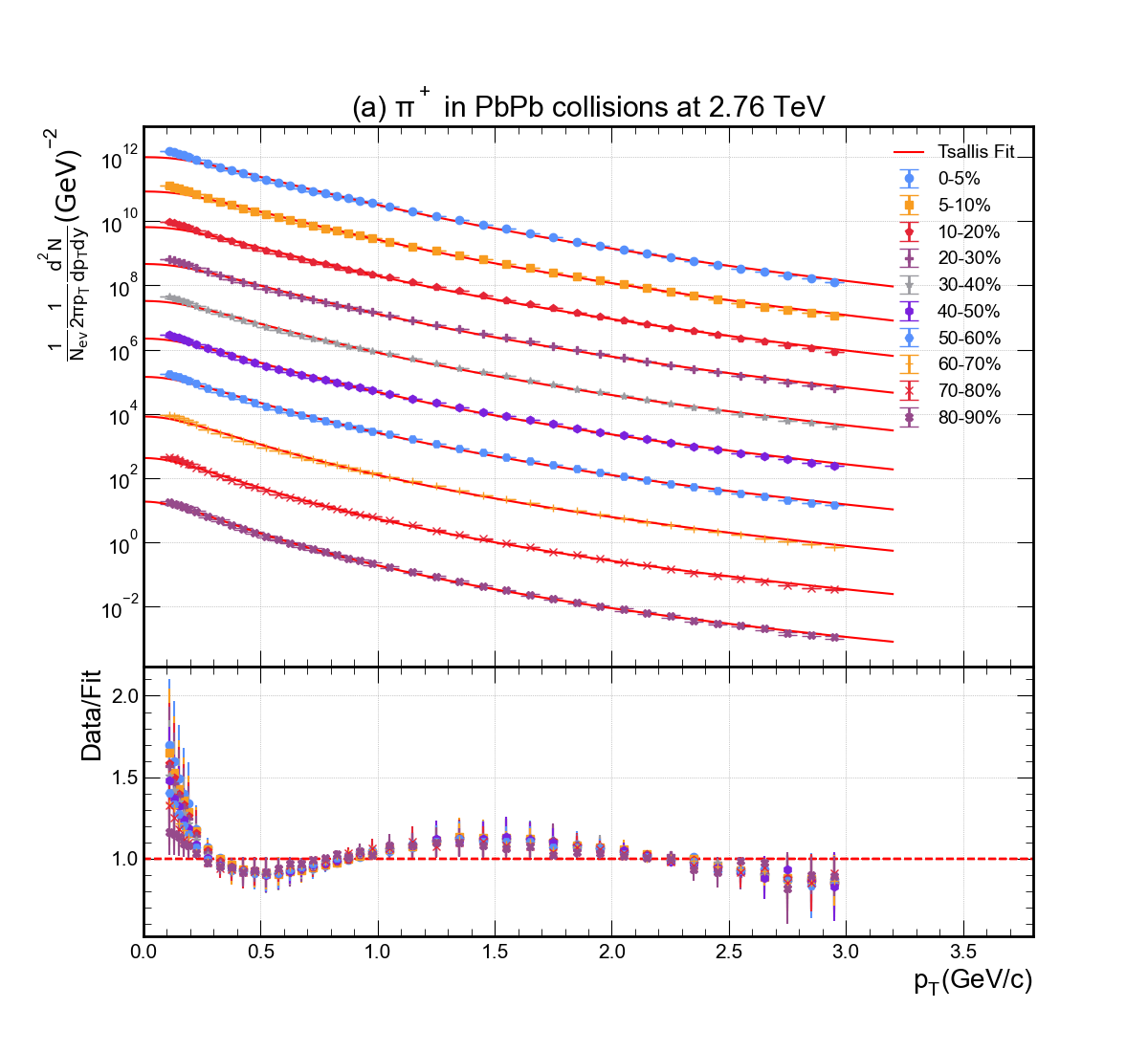} 
\includegraphics[width=0.48\textwidth]{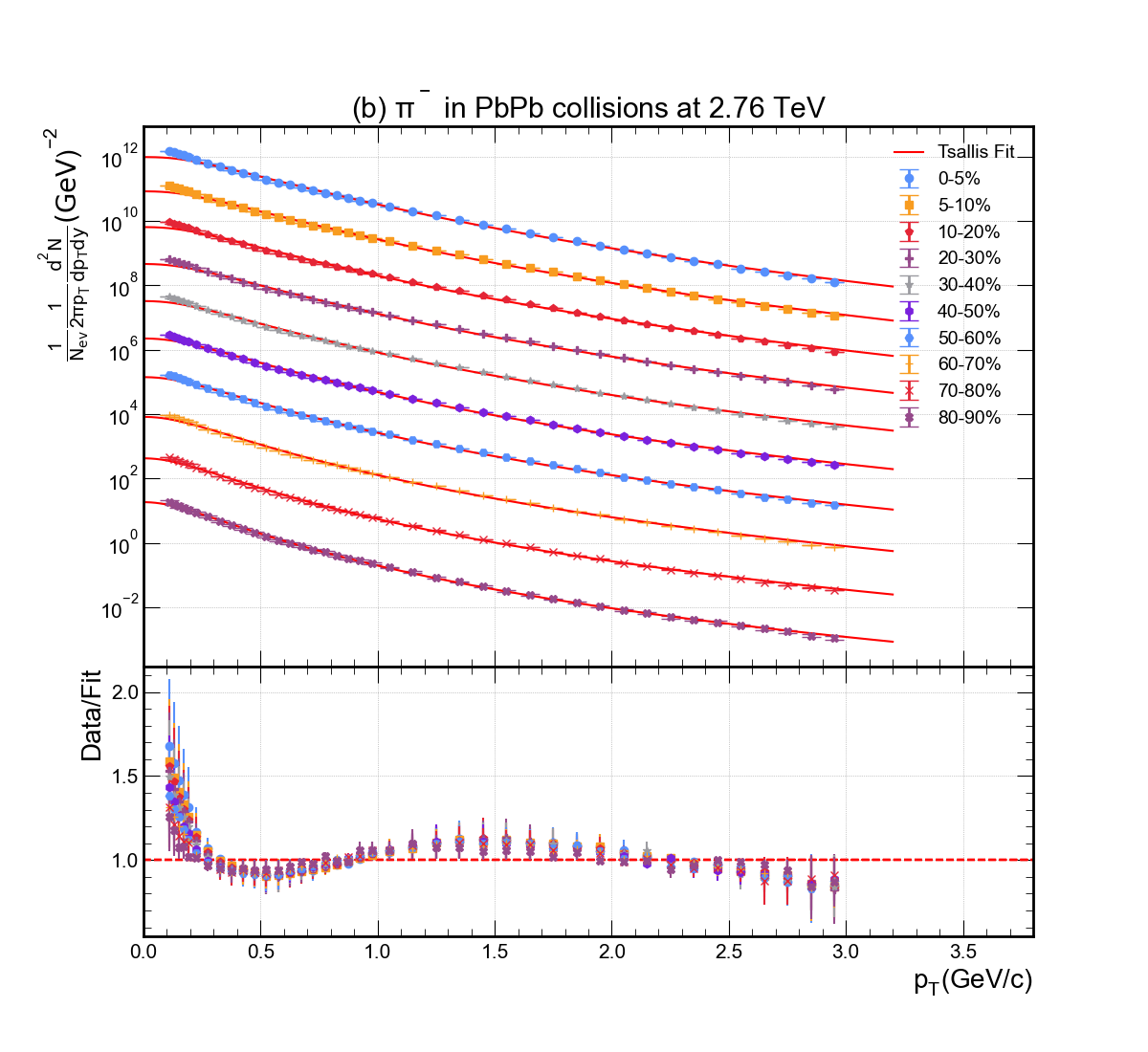} \vspace{-0.4cm}
\includegraphics[width=0.48\textwidth]{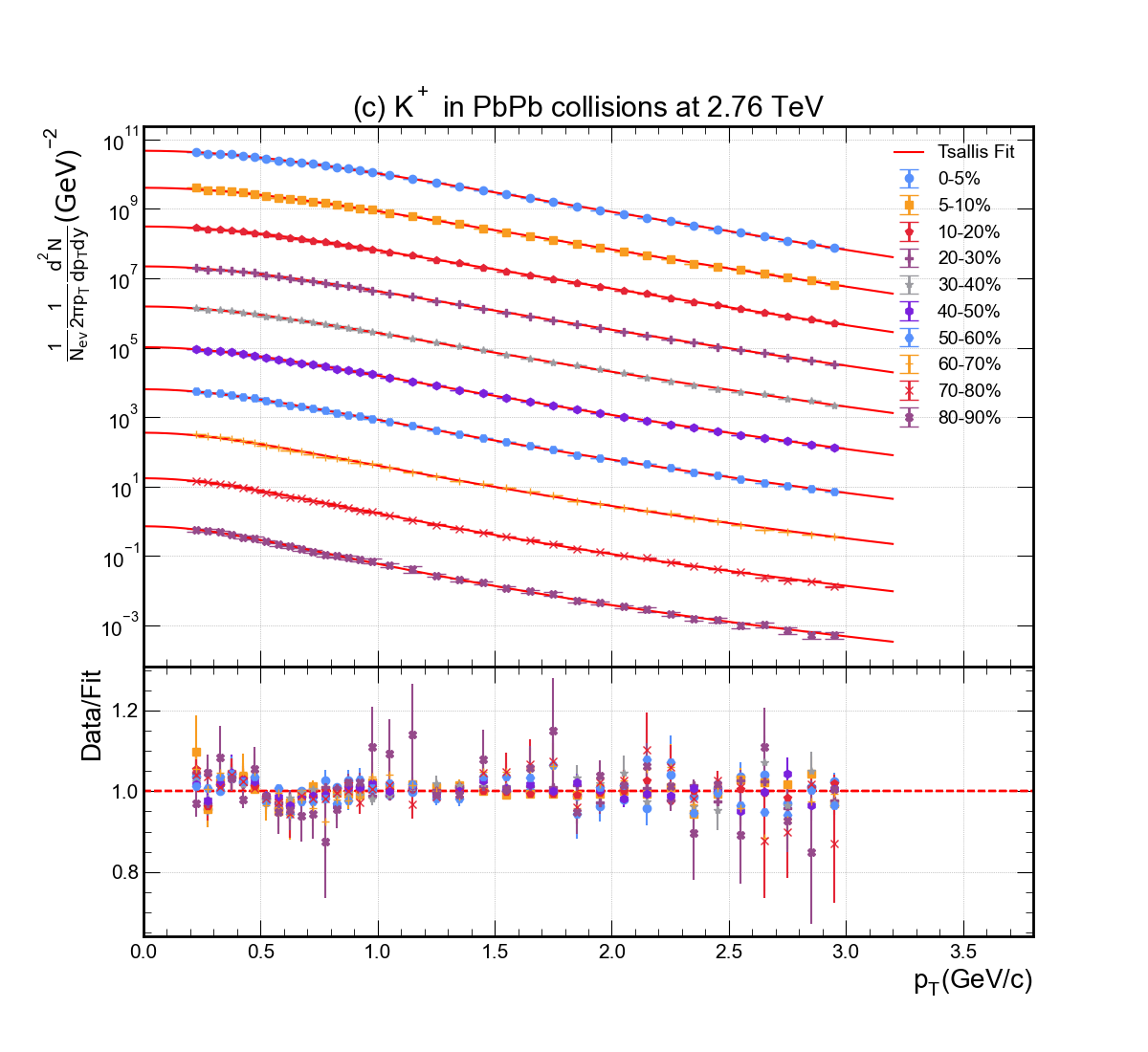}
\includegraphics[width=0.48\textwidth]{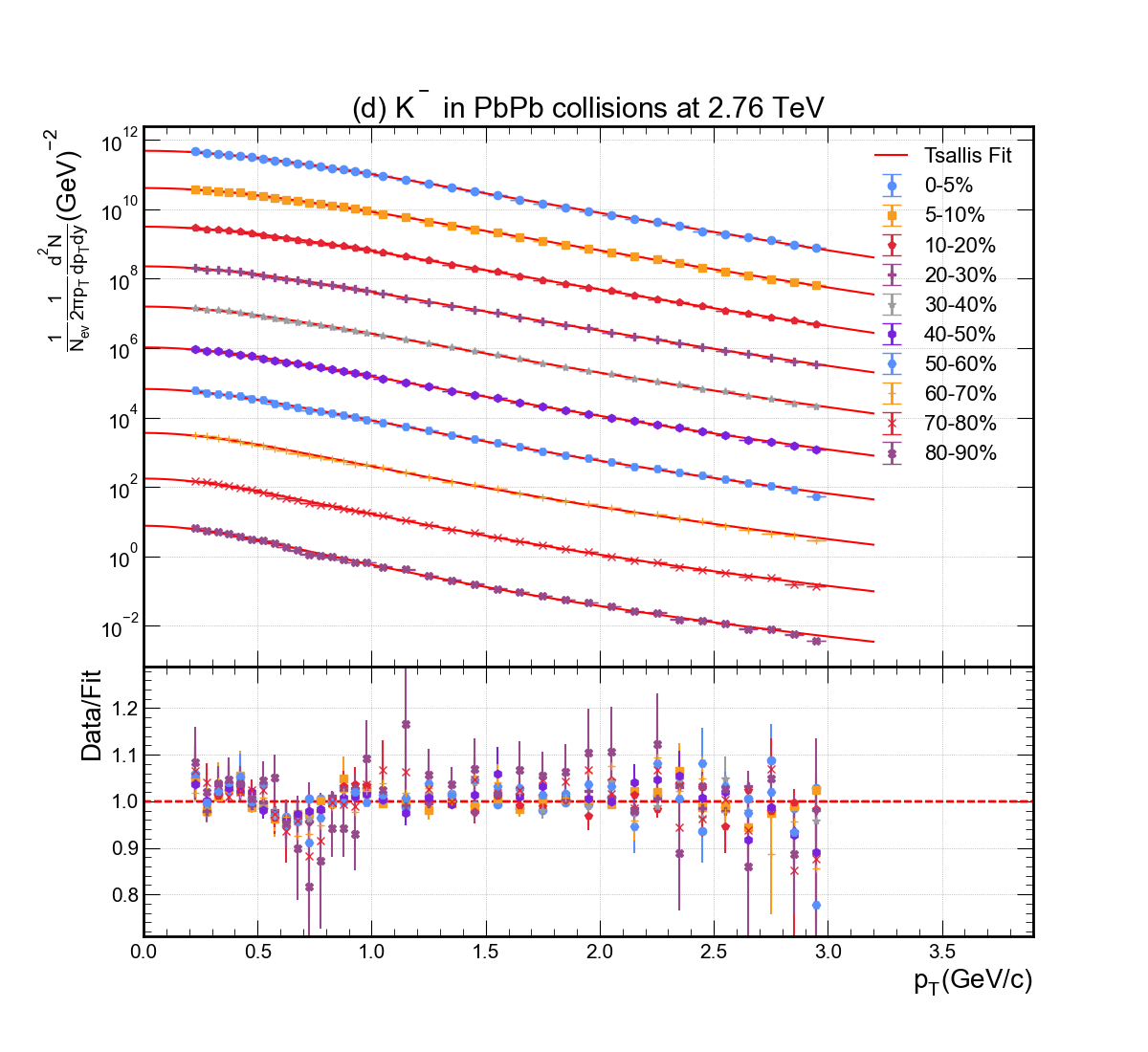} \vspace{-0.4cm}
\includegraphics[width=0.48\textwidth]{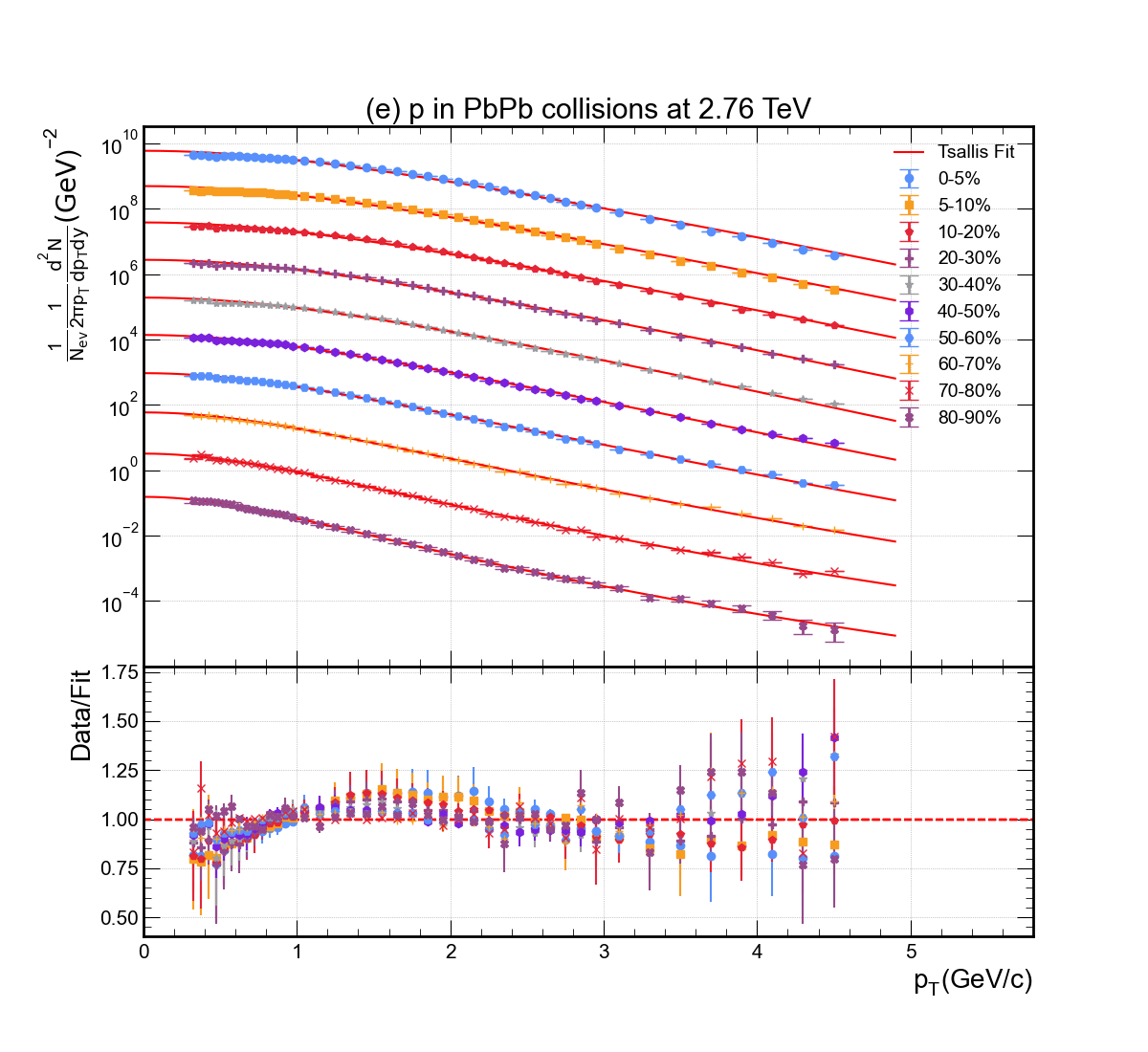}
\includegraphics[width=0.48\textwidth]{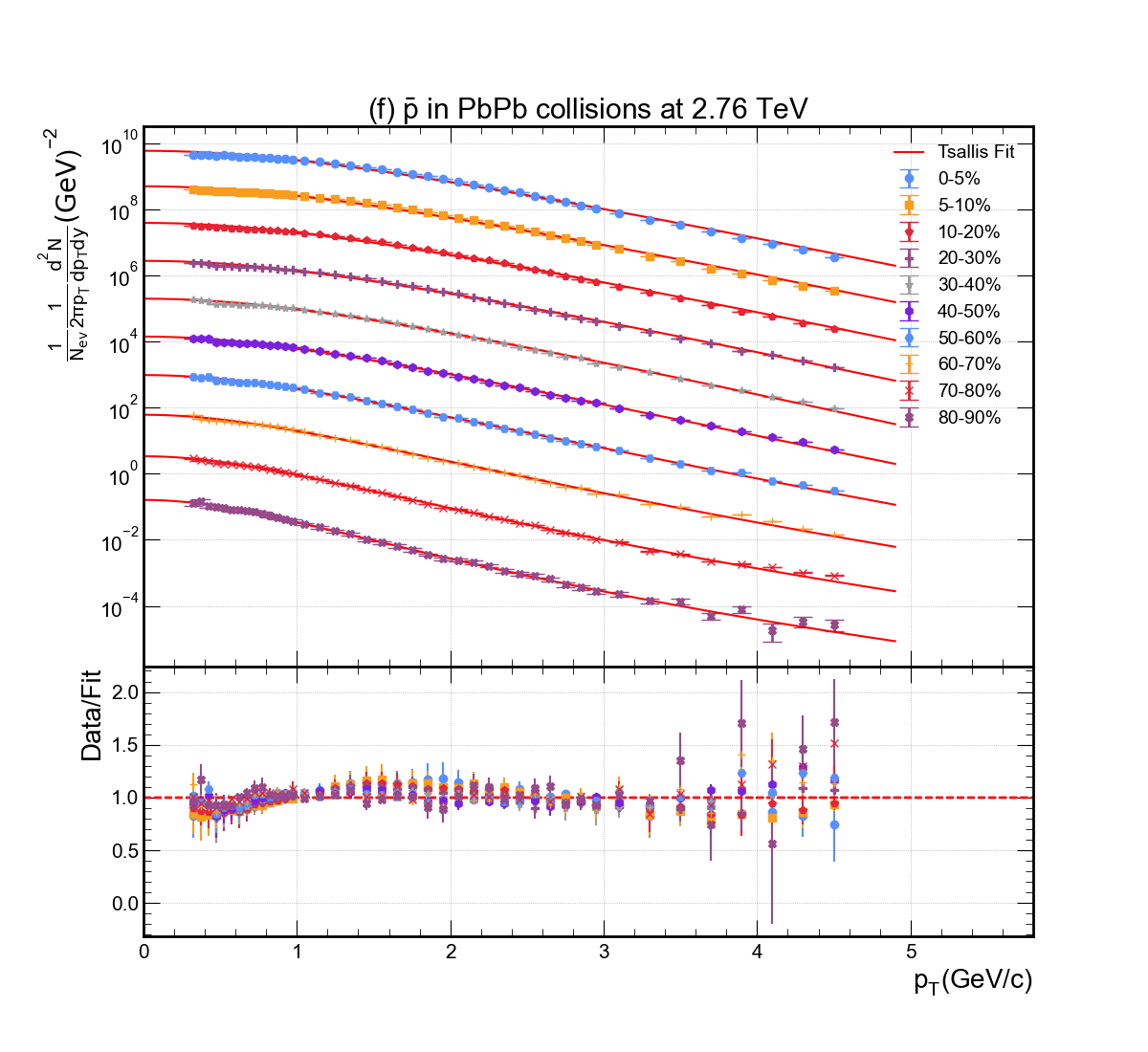}
\end{center}
\vspace{-0.4cm}
\caption{The $p_T$ distributions of $\pi^+$, $\pi^-$, $K^+$, $K^-$, $p$, and $\bar{p}$ are superimposed over the fit function across different centrality classes in Pb--Pb collisions at a collision energy of 2.76 TeV. Experimental data for various centrality values are shown with different coloured data points, while the solid lines give the fitting result, using Eq. (\ref{eqTsallisConsistent}). Each graph has been provided with a Data/Fit panel attached to the lower part.}
\end{figure}

\begin{figure}[t!]
\begin{center}
\hskip-0.153cm
\includegraphics[width=0.48\textwidth]{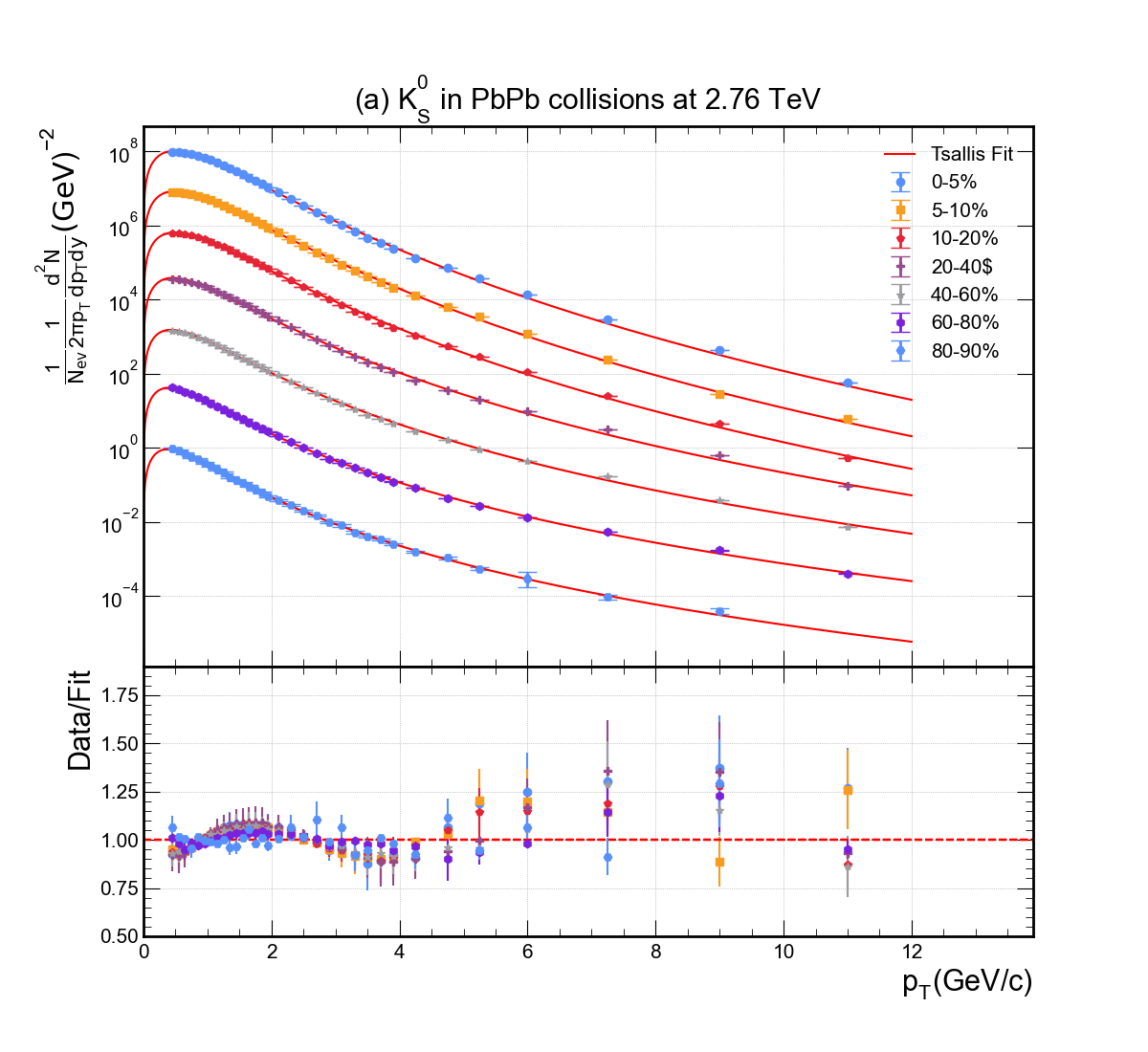}
\includegraphics[width=0.48\textwidth]{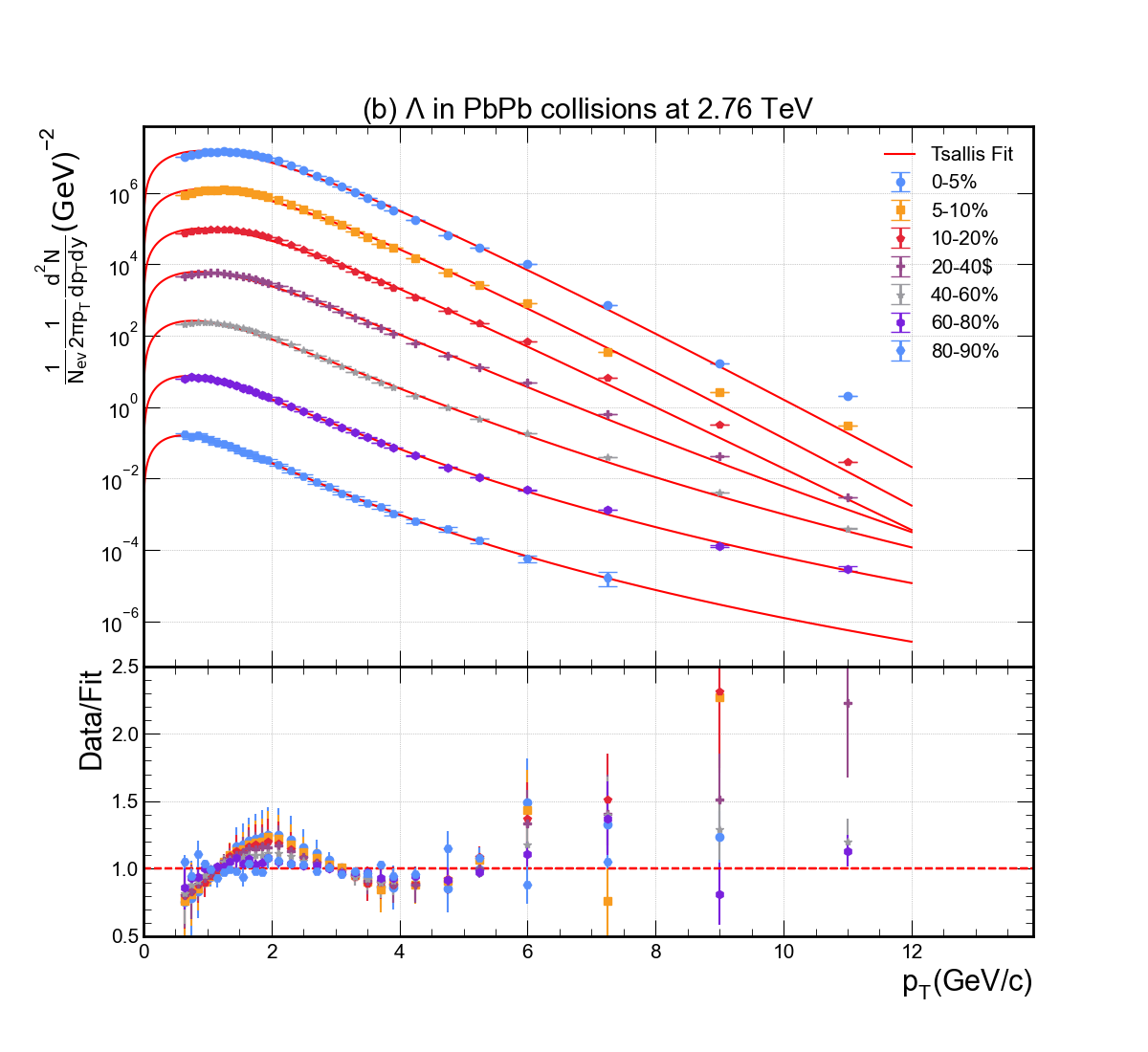}
\end{center}
\caption{The outcome of the fit function on the $p_T$ distribution of $K_s^0$ and $\Lambda$ measured by ALICE collaboration at 2.76 TeV of Pb--Pb collisions in different centrality classes. Experimental data for various centrality values are shown with different coloured data points, while the solid lines represent the fitting result obtained through Eq. (\ref{eqTsallisConsistent}). Each graph has been provided with a Data/Fit panel attached to the lower part.}
\end{figure}

The experimental data points are depicted with distinct symbols/colors for each centrality bin, while the solid lines represent the fitted Tsallis function. Each panel includes a data/fit ratio inset, which confirms that the deviations of the fit from the data are minimal and randomly distributed around unity, indicating a high-quality fit. The $\chi^2$/NDF values obtained listed in Table \ref{tab:tsallis}, quantitatively demonstrating the goodness of fit. 
Here, NDF denotes the number of degrees of freedom, and the parameter $\chi^2$ is defined as given in Ref. \cite{bashir2017centrality}:
\begin{align}
\chi^2=\sum_{i}\frac{(R_i^{\text{Exp}}-R_i^{\text{Theor}})^2}{\epsilon^2}.
\end{align}
In the above equation, $R_i^{\text{Exp}}$ represents the experimental data, $\epsilon$ is the measurement uncertainty, and $R_i^{\text{Theor}}$ gives the values calculated by our model.

\begin{scriptsize} 
\begin{longtable}{L{1.2cm} L{1.7cm} c c c c c c}
\caption{The table shows the values of the parameters including $T$, $q$, the normalization constant, $\chi^2$ and NDF obtained through the fitting procedure of the experimental data with Tsallis function given in Eq.\ 2.}\label{tab:tsallis}\\
\hline\hline
Particle Type & centrality & $N$ & $q$ & $T$ [GeV] & $\langle p_T \rangle$ [GeV] & $\chi^2$ & NDF \\
\hline
\endfirsthead

\multicolumn{8}{l}{\textit{Table \thetable\ (continued)}}\\
\hline\hline
Particle Type & centrality & $N$ & $q$ & $T$ [GeV] & $\langle p_T \rangle$ [GeV] & $\chi^2$ & NDF \\
\hline
\endhead

\hline
\multicolumn{8}{r}{\textit{Continued on next page}}\\
\endfoot

\hline\hline
\endlastfoot
\\
 & 00-05\% & 24314 $\pm$ 2929 & 1.1180 $\pm$ 0.0047 & 0.1173 $\pm$ 0.0044 & 0.2513 $\pm$ 0.0133 & 87.9 & 39 \\
 & 05-10\% & 21296 $\pm$ 2504 & 1.1214 $\pm$ 0.0046 & 0.1149 $\pm$ 0.0042 & 0.2480 $\pm$ 0.0132 & 89.9 & 39 \\
 & 10-20\% & 17091 $\pm$ 2002 & 1.1253 $\pm$ 0.0045 & 0.1122 $\pm$ 0.0041 & 0.2427 $\pm$ 0.0132 & 91.1 & 39 \\
 & 20-30\% & 12830 $\pm$ 1535 & 1.1310 $\pm$ 0.0046 & 0.1078 $\pm$ 0.0040 & 0.2354 $\pm$ 0.0133 & 93.7 & 39 \\
$\pi^+$ & 30-40\% & 9709 $\pm$ 1150 & 1.1367 $\pm$ 0.0044 & 0.1023 $\pm$ 0.0037 & 0.2243 $\pm$ 0.0130 & 90.3 & 39 \\
 & 40-50\% & 7104 $\pm$ 829 & 1.1420 $\pm$ 0.0042 & 0.0966 $\pm$ 0.0035 & 0.2136 $\pm$ 0.0129 & 86.1 & 39 \\
 & 50-60\% & 5078 $\pm$ 577 & 1.1484 $\pm$ 0.0039 & 0.0897 $\pm$ 0.0031 & 0.1996 $\pm$ 0.0132 & 77.8 & 39 \\
 & 60-70\% & 3343 $\pm$ 364 & 1.1552 $\pm$ 0.0036 & 0.0829 $\pm$ 0.0027 & 0.1860 $\pm$ 0.0131 & 67.7 & 39 \\
 & 70-80\% & 1918 $\pm$ 201 & 1.1605 $\pm$ 0.0034 & 0.0771 $\pm$ 0.0025 & 0.1744 $\pm$ 0.0128 & 58.6 & 39 \\
 & 80-90\% & 938 $\pm$ 83 & 1.1607 $\pm$ 0.0027 & 0.0723 $\pm$ 0.0019 & 0.1634 $\pm$ 0.0125 & 40.0 & 39 \\
 \\
\hline
\\
 & 00-05\% & 23514 $\pm$ 2723 & 1.1163 $\pm$ 0.0045 & 0.1188 $\pm$ 0.0042 & 0.2550 $\pm$ 0.0134 & 86.7 & 39 \\
 & 05-10\% & 20895 $\pm$ 2312 & 1.1203 $\pm$ 0.0043 & 0.1159 $\pm$ 0.0039 & 0.2498 $\pm$ 0.0133 & 81.8 & 39 \\
 & 10-20\% & 16635 $\pm$ 1848 & 1.1242 $\pm$ 0.0043 & 0.1133 $\pm$ 0.0039 & 0.2446 $\pm$ 0.0133 & 84.1 & 39 \\
 & 20-30\% & 12514 $\pm$ 1399 & 1.1296 $\pm$ 0.0043 & 0.1088 $\pm$ 0.0038 & 0.2372 $\pm$ 0.0132 & 84.8 & 39 \\
$\pi^-$ & 30-40\% & 9432 $\pm$ 1046 & 1.1355 $\pm$ 0.0042 & 0.1034 $\pm$ 0.0036 & 0.2258 $\pm$ 0.0131 & 82.1 & 39 \\
 & 40-50\% & 7177 $\pm$ 750 & 1.1433 $\pm$ 0.0038 & 0.0962 $\pm$ 0.0031 & 0.2130 $\pm$ 0.0130 & 70.0 & 39 \\
 & 50-60\% & 4919 $\pm$ 511 & 1.1477 $\pm$ 0.0037 & 0.0906 $\pm$ 0.0029 & 0.2017 $\pm$ 0.0129 & 67.5 & 39 \\
 & 60-70\% & 3215 $\pm$ 317 & 1.1540 $\pm$ 0.0033 & 0.0840 $\pm$ 0.0025 & 0.1883 $\pm$ 0.0129 & 57.9 & 39 \\
 & 70-80\% & 1879 $\pm$ 182 & 1.1598 $\pm$ 0.0031 & 0.0777 $\pm$ 0.0023 & 0.1754 $\pm$ 0.0128 & 51.6 & 39 \\
 & 80-90\% & 909 $\pm$ 73 & 1.1612 $\pm$ 0.0025 & 0.0728 $\pm$ 0.0017 & 0.1646 $\pm$ 0.0126 & 32.8 & 39 \\
 \\
\hline
 \\
 & 00-05\% & 715 $\pm$ 20 & 1.0412 $\pm$ 0.0020 & 0.2459 $\pm$ 0.0024 & 0.4738 $\pm$ 0.0201 & 2.3 & 34 \\
 & 05-10\% & 662 $\pm$ 22 & 1.0491 $\pm$ 0.0023 & 0.2363 $\pm$ 0.0027 & 0.4598 $\pm$ 0.0202 & 3.2 & 34 \\
 & 10-20\% & 530 $\pm$ 14 & 1.0533 $\pm$ 0.0018 & 0.2310 $\pm$ 0.0021 & 0.4527 $\pm$ 0.0198 & 1.9 & 34 \\
 & 20-30\% & 421 $\pm$ 13 & 1.0622 $\pm$ 0.0019 & 0.2183 $\pm$ 0.0022 & 0.4333 $\pm$ 0.0196 & 2.2 & 34 \\
 & 30-40\% & 352 $\pm$ 12 & 1.0746 $\pm$ 0.0020 & 0.2006 $\pm$ 0.0023 & 0.4071 $\pm$ 0.0196 & 2.5 & 34 \\
$K^+$ & 40-50\% & 298 $\pm$ 11 & 1.0874 $\pm$ 0.0019 & 0.1811 $\pm$ 0.0021 & 0.3762 $\pm$ 0.0192 & 2.3 & 34 \\
 & 50-60\% & 239 $\pm$ 15 & 1.1002 $\pm$ 0.0031 & 0.1617 $\pm$ 0.0033 & 0.3459 $\pm$ 0.0188 & 5.4 & 34 \\
 & 60-70\% & 193 $\pm$ 18 & 1.1152 $\pm$ 0.0038 & 0.1401 $\pm$ 0.0039 & 0.3100 $\pm$ 0.0188 & 8.0 & 34 \\
 & 70-80\% & 134 $\pm$ 13 & 1.1265 $\pm$ 0.0034 & 0.1233 $\pm$ 0.0036 & 0.2816 $\pm$ 0.0185 & 6.3 & 34 \\
 & 80-90\% & 86 $\pm$ 16 & 1.1373 $\pm$ 0.0055 & 0.1061 $\pm$ 0.0056 & 0.2526 $\pm$ 0.0185 & 12.3 & 34 \\
 \\
\hline
\\
 & 00-05\% & 799 $\pm$ 26 & 1.0477 $\pm$ 0.0022 & 0.2362 $\pm$ 0.0026 & 0.4589 $\pm$ 0.0201 & 2.5 & 34 \\
 & 05-10\% & 699 $\pm$ 25 & 1.0512 $\pm$ 0.0026 & 0.2320 $\pm$ 0.0029 & 0.4536 $\pm$ 0.0202 & 3.4 & 34 \\
 & 10-20\% & 570 $\pm$ 22 & 1.0566 $\pm$ 0.0026 & 0.2251 $\pm$ 0.0029 & 0.4437 $\pm$ 0.0200 & 3.7 & 34 \\
 & 20-30\% & 478 $\pm$ 19 & 1.0692 $\pm$ 0.0024 & 0.2089 $\pm$ 0.0027 & 0.4200 $\pm$ 0.0198 & 3.2 & 34 \\
$K^-$ & 30-40\% & 392 $\pm$ 18 & 1.0796 $\pm$ 0.0027 & 0.1934 $\pm$ 0.0029 & 0.3954 $\pm$ 0.0197 & 3.6 & 34 \\
 & 40-50\% & 328 $\pm$ 16 & 1.0919 $\pm$ 0.0026 & 0.1749 $\pm$ 0.0028 & 0.3669 $\pm$ 0.0195 & 3.5 & 34 \\
 & 50-60\% & 293 $\pm$ 26 & 1.1065 $\pm$ 0.0041 & 0.1521 $\pm$ 0.0042 & 0.3297 $\pm$ 0.0195 & 8.8 & 34 \\
 & 60-70\% & 212 $\pm$ 21 & 1.1174 $\pm$ 0.0039 & 0.1362 $\pm$ 0.0041 & 0.3032 $\pm$ 0.0194 & 7.9 & 34 \\
 & 70-80\% & 150 $\pm$ 20 & 1.1312 $\pm$ 0.0047 & 0.1186 $\pm$ 0.0047 & 0.2753 $\pm$ 0.0195 & 9.8 & 34 \\
 & 80-90\% & 129 $\pm$ 33 & 1.1466 $\pm$ 0.0069 & 0.0948 $\pm$ 0.0069 & 0.2338 $\pm$ 0.0193 & 17.7 & 34 \\
 \\
 \\
\hline
\\
 & 00-05\% & 60 $\pm$ 5 & 1.0003 $\pm$ 0.0044 & 0.4185 $\pm$ 0.0088 & 0.7840 $\pm$ 0.0252 & 73.9 & 40 \\
 & 05-10\% & 50 $\pm$ 3 & 1.0004 $\pm$ 0.0044 & 0.4168 $\pm$ 0.0080 & 0.7813 $\pm$ 0.0251 & 67.8 & 40 \\
 & 10-20\% & 39 $\pm$ 4 & 1.0046 $\pm$ 0.0067 & 0.4136 $\pm$ 0.0134 & 0.7761 $\pm$ 0.0250 & 53.4 & 40 \\
 & 20-30\% & 30 $\pm$ 2 & 1.0047 $\pm$ 0.0054 & 0.4039 $\pm$ 0.0104 & 0.7604 $\pm$ 0.0252 & 35.9 & 40 \\
$p$ & 30-40\% & 22 $\pm$ 2 & 1.0153 $\pm$ 0.0046 & 0.3914 $\pm$ 0.0089 & 0.7488 $\pm$ 0.0250 & 24.2 & 40 \\
 & 40-50\% & 22 $\pm$ 2 & 1.0181 $\pm$ 0.0039 & 0.3419 $\pm$ 0.0074 & 0.6693 $\pm$ 0.0249 & 17.8 & 40 \\
 & 50-60\% & 23 $\pm$ 2 & 1.0353 $\pm$ 0.0037 & 0.2942 $\pm$ 0.0067 & 0.5987 $\pm$ 0.0248 & 14.5 & 40 \\
 & 60-70\% & 28 $\pm$ 3 & 1.0586 $\pm$ 0.0037 & 0.2362 $\pm$ 0.0064 & 0.5116 $\pm$ 0.0248 & 14.4 & 40 \\
 & 70-80\% & 32 $\pm$ 5 & 1.0775 $\pm$ 0.0044 & 0.1891 $\pm$ 0.0073 & 0.4381 $\pm$ 0.0245 & 17.9 & 40 \\
 & 80-90\% & 42 $\pm$ 8 & 1.0909 $\pm$ 0.0041 & 0.1452 $\pm$ 0.0066 & 0.3661 $\pm$ 0.0242 & 10.7 & 40 \\
 \\
\hline
\\
 & 00-05\% & 58 $\pm$ 6 & 1.0008 $\pm$ 0.0070 & 0.4187 $\pm$ 0.0138 & 0.7846 $\pm$ 0.0249 & 61.5 & 40 \\
 & 05-10\% & 50 $\pm$ 5 & 1.0023 $\pm$ 0.0075 & 0.4161 $\pm$ 0.0149 & 0.7812 $\pm$ 0.0249 & 61.4 & 40 \\
 & 10-20\% & 40 $\pm$ 2 & 1.0026 $\pm$ 0.0038 & 0.4117 $\pm$ 0.0067 & 0.7744 $\pm$ 0.0248 & 46.8 & 40 \\
 & 20-30\% & 30 $\pm$ 2 & 1.0051 $\pm$ 0.0044 & 0.4044 $\pm$ 0.0079 & 0.7633 $\pm$ 0.0246 & 32.7 & 40 \\
$\bar p$ & 30-40\% & 23 $\pm$ 2 & 1.0149 $\pm$ 0.0047 & 0.3891 $\pm$ 0.0091 & 0.7447 $\pm$ 0.0245 & 23.3 & 40 \\
 & 40-50\% & 22 $\pm$ 2 & 1.0152 $\pm$ 0.0041 & 0.3462 $\pm$ 0.0078 & 0.6745 $\pm$ 0.0244 & 16.9 & 40 \\
 & 50-60\% & 24 $\pm$ 2 & 1.0350 $\pm$ 0.0039 & 0.2921 $\pm$ 0.0072 & 0.5955 $\pm$ 0.0240 & 14.1 & 40 \\
 & 60-70\% & 27 $\pm$ 3 & 1.0559 $\pm$ 0.0044 & 0.2387 $\pm$ 0.0076 & 0.5152 $\pm$ 0.0239 & 16.8 & 40 \\
 & 70-80\% & 33 $\pm$ 5 & 1.0753 $\pm$ 0.0041 & 0.1898 $\pm$ 0.0069 & 0.4388 $\pm$ 0.0239 & 13.1 & 40 \\
 & 80-90\% & 55 $\pm$ 17 & 1.0937 $\pm$ 0.0061 & 0.1377 $\pm$ 0.0098 & 0.3535 $\pm$ 0.0236 & 22.3 & 40 \\
 \\
\hline
\\
 & 00-05\% & 8553 $\pm$ 849 & 1.0722 $\pm$ 0.0025 & 0.1997 $\pm$ 0.0049 & 0.4049 $\pm$ 0.0205 & 65.1 & 31 \\
 & 05-10\% & 6988 $\pm$ 611 & 1.0740 $\pm$ 0.0022 & 0.1994 $\pm$ 0.0044 & 0.4043 $\pm$ 0.0204 & 49.3 & 31 \\
 & 10-20\% & 5851 $\pm$ 492 & 1.0798 $\pm$ 0.0021 & 0.1929 $\pm$ 0.0040 & 0.3955 $\pm$ 0.0204 & 43.8 & 31 \\
$K^0_S$ & 20-40\% & 5163 $\pm$ 635 & 1.0973 $\pm$ 0.0025 & 0.1679 $\pm$ 0.0050 & 0.3561 $\pm$ 0.0203 & 79.5 & 31 \\
 & 40-60\% & 3250 $\pm$ 365 & 1.1139 $\pm$ 0.0019 & 0.1437 $\pm$ 0.0038 & 0.3167 $\pm$ 0.0198 & 48.9 & 31 \\
 & 60-80\% & 1624 $\pm$ 159 & 1.1308 $\pm$ 0.0014 & 0.1186 $\pm$ 0.0027 & 0.2749 $\pm$ 0.0196 & 22.6 & 31 \\
 & 80-90\% & 490 $\pm$ 62 & 1.1359 $\pm$ 0.0019 & 0.1081 $\pm$ 0.0032 & 0.2527 $\pm$ 0.0194 & 15.7 & 30 \\
 \\
\hline
\\
 & 00-05\% & 344 $\pm$ 29 & 1.0012 $\pm$ 0.0029 & 0.4229 $\pm$ 0.0074 & 0.8111 $\pm$ 0.0405 & 88.5 & 29 \\
 & 05-10\% & 292 $\pm$ 43 & 1.0028 $\pm$ 0.0058 & 0.4226 $\pm$ 0.0163 & 0.8106 $\pm$ 0.0405 & 68.9 & 29 \\
 & 10-20\% & 252 $\pm$ 36 & 1.0047 $\pm$ 0.0054 & 0.4103 $\pm$ 0.0152 & 0.7936 $\pm$ 0.0403 & 55.4 & 29 \\
$\Lambda$ & 20-40\% & 241 $\pm$ 38 & 1.0236 $\pm$ 0.0049 & 0.3569 $\pm$ 0.0137 & 0.7176 $\pm$ 0.0405 & 50.3 & 29 \\
 & 40-60\% & 248 $\pm$ 40 & 1.0508 $\pm$ 0.0038 & 0.2765 $\pm$ 0.0102 & 0.5979 $\pm$ 0.0392 & 31.3 & 29 \\
 & 60-80\% & 240 $\pm$ 41 & 1.0771 $\pm$ 0.0029 & 0.2014 $\pm$ 0.0073 & 0.4807 $\pm$ 0.0389 & 15.5 & 29 \\
 & 80-90\% & 164 $\pm$ 34 & 1.0894 $\pm$ 0.0029 & 0.1558 $\pm$ 0.0066 & 0.4051 $\pm$ 0.0382 & 6.5 & 27 \\
 \\
\end{longtable}
\end{scriptsize}

The ability of a single Tsallis function to describe the spectra over the entire measured $p_T$ range, from low-$p_T$ thermal region to the high-$p_T$ hard scattering tail, underscores the versatility of this statistical approach. Indeed, previous studies have also found that Tsallis distributions can accurately reproduce hadron $p_T$ spectra in high-energy collisions over wide momentum ranges, owing to their power-law asymptotic behavior (controlled by the non-extensive parameter $q$) combined with an exponential-like attenuation at low $p_T$ (governed by the effective temperature $T$) \cite{Cleymans2012, CLEYMANS2013351, buzatu}.

The Tsallis function provides a quantitatively reasonable description of the spectra across centralities and particle species, with $\chi^2/NDF$ values mostly of order unity and only moderate variations with centrality and particle type. In central Pb–Pb collisions, the spectra are relatively flatter and harder, especially for heavier particles, reflecting the stronger collective flow and higher effective temperature of the system. In contrast, in peripheral collisions, the spectra fall off more steeply with $p_T$, indicating a cooler and less explosive source \cite{source1, hydroModels}. The Tsallis fit captures both regimes by yielding larger $T$ parameters for central collisions and smaller $T$ for peripheral ones details quantified below. Additionally, the success of the fits for the multi-strange baryon $\Lambda$ and the $K^0_S$ meson Fig.~2 is particularly encouraging. These hadrons involve different production mechanisms e.g., strangeness conservation and baryon number transport, compared to pions and kaons, yet the same functional form describes their $p_T$ distributions well. This implies that the Tsallis distribution, being a purely statistical parametrization, is sufficiently general to account for a variety of particle species without invoking separate model assumptions for each. We emphasize that no collective flow velocity or external physics assumptions are explicitly built into the Tsallis function; rather, effects such as radial flow are effectively encoded in the spectral shape. This is consistent with other analyses where Tsallis or QCD-inspired power-law functions were found to empirically fit identified particle spectra in heavy-ion and even small-system collisions \cite{yang2021dependence, badshah2023systematic}. Overall, the close agreement between data and fits in Figs.~1–2 establishes a solid foundation for extracting physical parameters from the Tsallis fits and comparing them across centralities and particle types.

\subsection{Centrality Dependence of Tsallis Parameters}
The centrality evolution of the Tsallis fit parameters are summarized in \ref{tab:tsallis} and are plotted in Fig.~~3. Panel~~a of Fig.~~3 shows the effective temperature $T$ as a function of collision centrality for each particle species. A clear trend is observed: $T$ rises monotonically from peripheral to central collisions. In the most peripheral bin e.g. 80–90\% centrality, $T$ is relatively low for all hadrons, indicating cooler effective source temperatures in small, less violent collisions. As collisions become more central, system size and participant nucleon number increase, the deposited energy density is higher, leading to higher excitation of the system and consequently higher $T$ parameters for the emitted particles. For example, in central 0–5\% Pb–Pb collisions at 2.76~~TeV, we extracted $T\approx0.12$~GeV for pions and up to $T\approx0.42$~~GeV for protons and $\Lambda$ hyperons. In peripheral 80–90\% collisions, the corresponding $T$ values are significantly smaller; see Table \ref{tab:tsallis} for detailed values. This behavior is consistent with the expectation that central heavy-ion collisions approach local thermal equilibrium more closely and generate hotter systems than peripheral collisions. Similar centrality-dependent increases in slope parameters have been reported in other studies; for instance, in Ref.~~\cite{yang2021dependence}, a rise in $T$ from peripheral to central events was also noted, reflecting greater energy transfer per nucleon in more central interactions. Our findings thus reinforce the notion that higher participant densities and more central collisions yield a higher effective temperature of the emitted hadrons, qualitatively consistent with hydrodynamic descriptions of heavy-ion collisions where more central collisions produce stronger collective expansion and higher kinetic temperatures \cite{collectivity, hydroModels}.

In addition to increasing with centrality, the Tsallis $T$ exhibits a systematic dependence on the particle rest mass, which provides evidence for a mass-dependent freeze-out scenario. At any given centrality, we find that heavier hadrons emerge with larger fitted $T$ values than lighter hadrons. For example, in central collisions $T_{\Lambda}\approx0.42$~GeV is much higher than $T_{\pi}\approx0.12$~GeV, with kaons and protons taking intermediate values. This hierarchy $T_{\mathrm{heavy}} > T_{\mathrm{light}}$ suggests that heavier particles have harder $p_T$ spectra flatter slope \cite{hydroModels} and effectively “freeze out” earlier at a higher kinetic temperature than lighter ones. Physically, particles with larger mass have greater inertia and receive a larger momentum kick from the collective expansion blue-shift effect, resulting in a higher apparent temperature of their spectra. Consequently, heavier hadrons decouple from the fireball sooner, when the system is still relatively hot, whereas lighter hadrons continue to interact and cool to lower temperatures before decoupling. This multiple freeze-out picture, in which different particle species decouple at different temperatures, is supported by our observation of distinct $T$ for pions, kaons, and protons. It aligns with the concept of “differential freeze-out” discussed in the literature \cite{badshah2023evolution, Aj}, where each hadron species can have its own kinetic freeze-out temperature. In contrast, a scenario with a single common freeze-out temperature for all particles is disfavored by our results unless one allows additional physics e.g. different flow velocities or non-thermal contributions to compensate for the mass dependence. We note that our fits were performed independently for each particle species rather than a combined simultaneous fit, which enables this multi-freeze-out behavior to manifest clearly. 
A combined fit imposing a single $T$ for all hadrons would be oversimplified given the diverse masses and could not reproduce the detailed spectral differences \cite{barnby2009systematic}. Our approach, treating each species separately, revealing the intrinsic mass-dependent effects that are washed out in one-size-fits-all models.
\begin{figure}[pt!]
\begin{center}
\hskip-0.153cm
\includegraphics[width=0.50\textwidth]{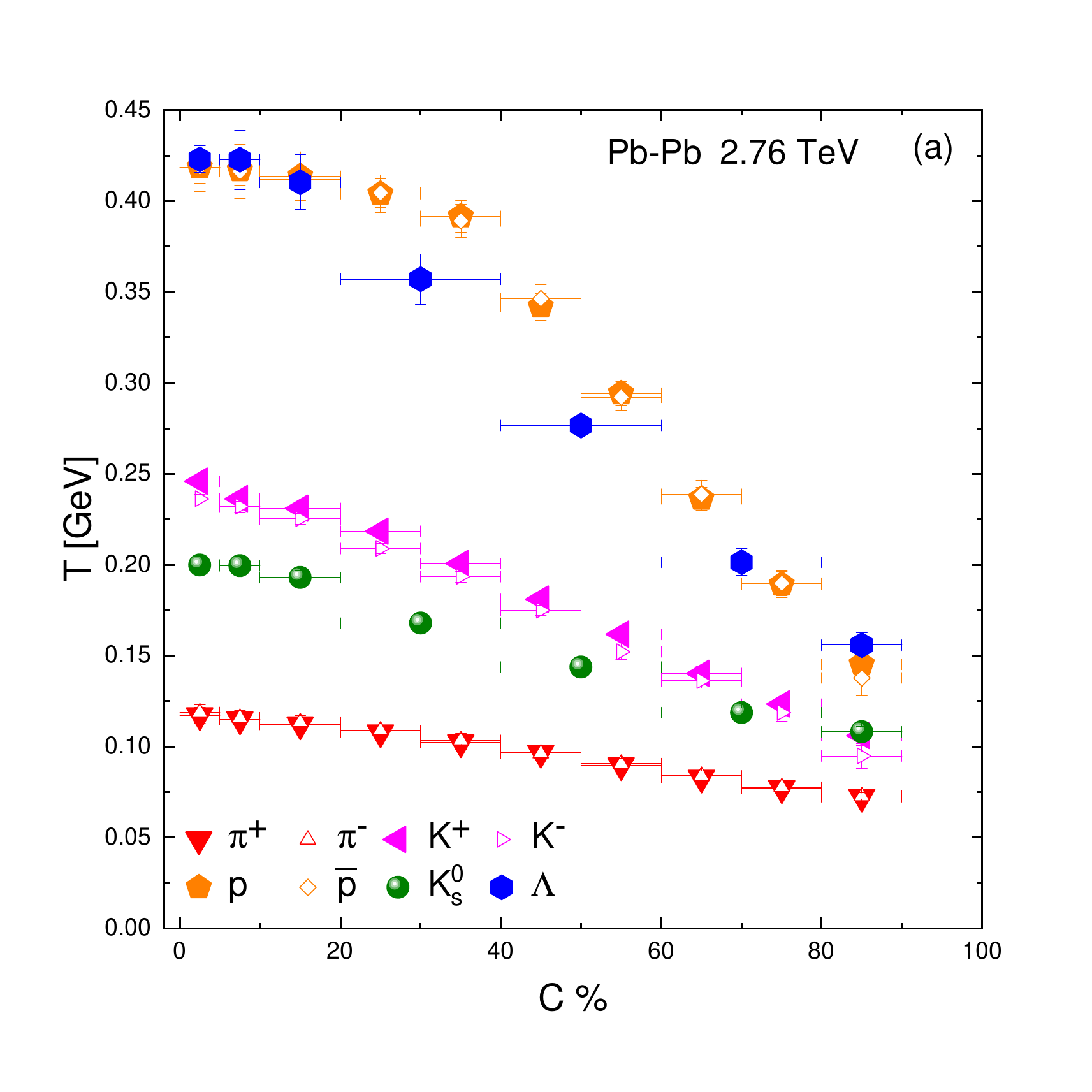}
\includegraphics[width=0.50\textwidth]{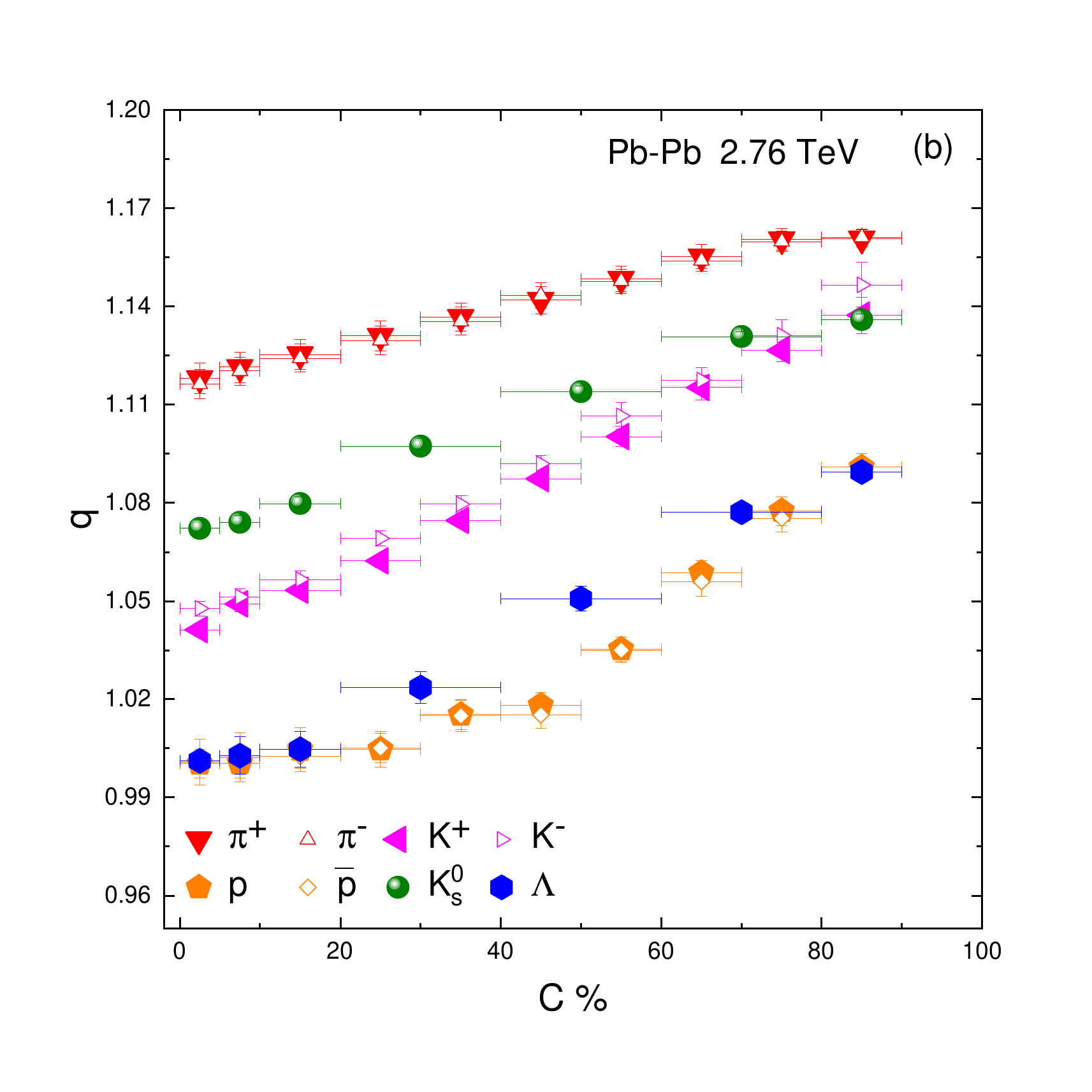}
\includegraphics[width=0.54\textwidth]{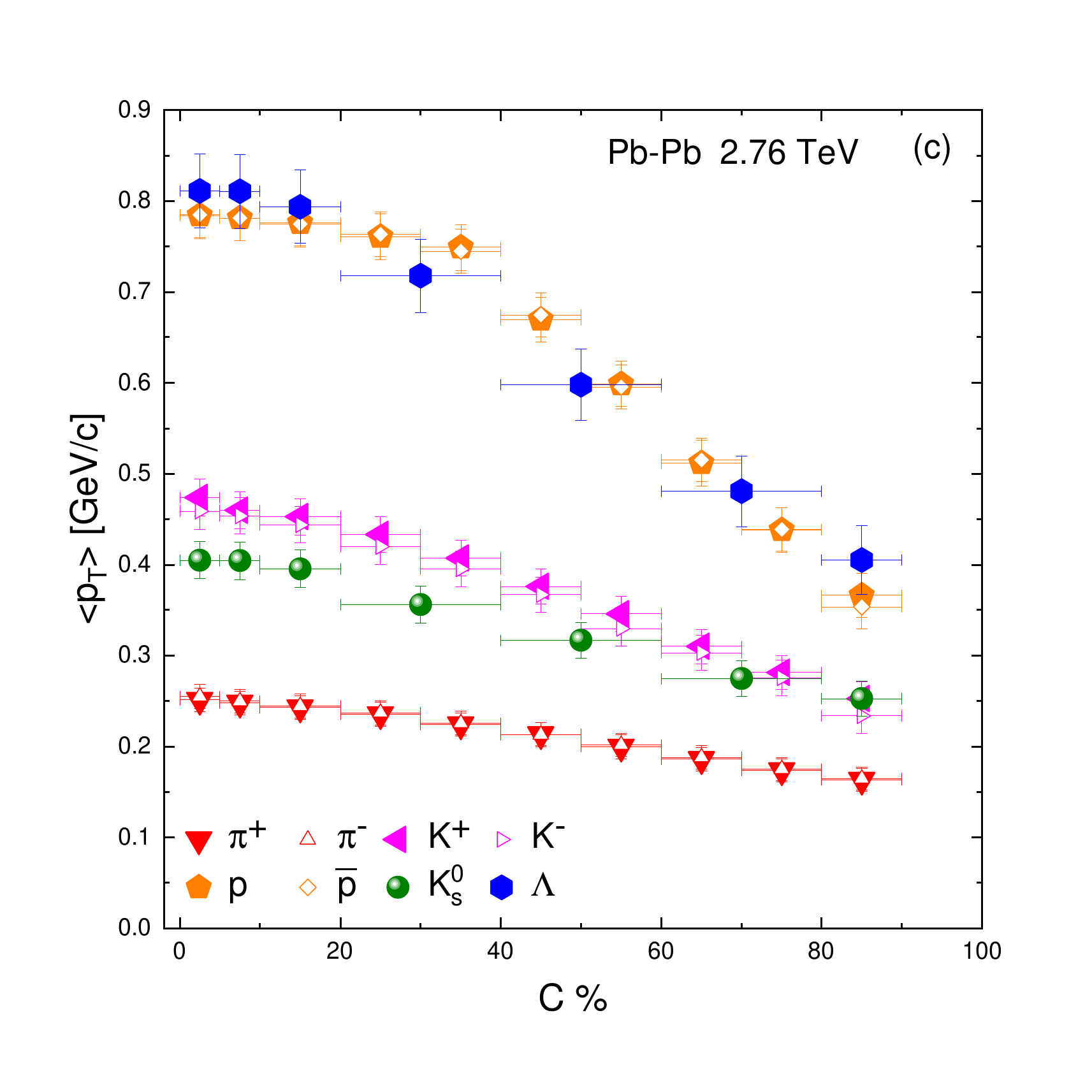}
\end{center}
\caption{The centrality dependence of (a) $T$, (b) $q$ and (c) $\langle p_T\rangle$.}
\end{figure}

Figure~~3b displays the non-extensivity parameter $q$ as a function of centrality. In contrast to $T$, the parameter $q$ decreases from peripheral to central collisions. Peripheral collisions yield $q$ values significantly above unity (e.g. $q\sim1.15$–1.20 for the most peripheral class), indicating strong deviations from equilibrium in these small, dilute systems. As the collisions become more central and the system density and particle multiplicity grow, $q$ approaches closer to 1. For the top centralities 0–5\% and 5–10\%, we obtain $q$ values only slightly above 1 (typically $q \approx 1.05$–1.08, varying by particle species), suggesting that the bulk matter created in central Pb–Pb collisions is much nearer to local thermal equilibrium. This trend makes intuitive sense: central collisions produce a higher density fireball with numerous rescatterings, which drive the system toward equilibrium (hence $q\to1$). Peripheral collisions, on the other hand, are more akin to “small systems” with fewer interactions, so the particle emission exhibits noticeable non-thermal tails (larger $q$). Our findings for $q(\text{centrality})$ are in qualitative agreement with earlier analyses. For example, Patra \emph{et~~al.} observed that at LHC energies $q$ is roughly constant or slightly decreases toward central collisions, and other works using Tsallis fits in heavy-ion collisions have also reported $q$ values closer to unity in central events \cite{yang2021dependence}. The values of $q$ extracted here (approximately 1.05–1.1 in central and up to $\sim1.2$ in peripheral) signify that even the most central heavy-ion collisions are not perfectly thermal ($q$ would equal 1 for an ideal Boltzmann-Gibbs equilibrium), but they are much more thermalized than peripheral collisions. It is also interesting to examine the species-dependence of $q$. We observe that for a fixed centrality, heavier particles tend to have $q$ values slightly closer to 1 compared to lighter particles see Table \ref{tab:tsallis}. In other words, the degree of non-equilibrium appears smaller for heavy hadrons, which could imply that heavier hadrons achieve thermalization more rapidly or are emitted from regions of the fireball that are closer to equilibrium. This pattern is consistent with the earlier notion of multi-freeze-out: heavier hadrons freezing out earlier might experience conditions closer to thermal equilibrium (hence lower $q$). This is because they decouple when the system is denser and interactions are still frequent. Meanwhile, lighter hadrons decouple later when the system is more dilute and beginning to fall out of equilibrium, yielding slightly higher $q$. These interpretations reinforce how $q$ serves as an \emph{entropy index} capturing the equilibration status of the system: lower $q$ closer to 1 indicates a system nearer to local thermal equilibrium, while higher $q$ reflects a departure from equilibrium due to insufficient rescattering or fragmentary sources of particle production.

Finally, Fig.~~3c presents the mean transverse momentum $\langle p_T\rangle$ of emitted particles as a function of centrality. The $\langle p_T\rangle$ values were calculated using Eq. (7) for each particle species and centrality bin and are also listed in Table \ref{tab:tsallis}. We find that $\langle p_T\rangle$ increases appreciably from peripheral to central collisions, mirroring the trend observed for $T$. In central Pb–Pb events, all particles have a higher average $p_T$ than in peripheral events. For instance, pions in central collisions have $\langle p_T\rangle$ on the order of a few hundred MeV/$c$, larger by roughly 30–50\% compared to pions from peripheral collisions. This systematic increase of $\langle p_T\rangle$ with centrality is a well-known signature of stronger radial flow in more central collisions. As the fireball volume and participant number grow, the collective expansion becomes more pronounced, imparting larger transverse momenta to the particles on average. Additionally, $\langle p_T\rangle$ depends on the particle mass: at a given centrality, heavier particles emerge with higher average $p_T$ than lighter ones (e.g. $\langle p_T\rangle_p > \langle p_T\rangle_{\pi}$). This mass ordering of $\langle p_T\rangle$ is again a hallmark of radial flow: heavier hadrons pick up a larger momentum boost from the common flow velocity field, leading to higher $\langle p_T\rangle$ \cite{hydroModels, Andronic2011}. Our results in Fig.~3c quantitatively corroborate this behavior. For example, in mid-central collisions around 30–40\%, we find $\langle p_T\rangle_\Lambda$ to be significantly larger than $\langle p_T\rangle_{\pi}$, which is consistent with expectations from hydrodynamic models where a single transverse flow velocity would give $p_T \approx m_T \sinh\rho$ (with $\rho$ the flow rapidity) so that more massive particles attain higher momentum.

\subsection{Correlations Among Tsallis Parameters and Freeze-Out Extraction}
While examining individual trends of $T$, $q$, and $\langle p_T\rangle$ is informative, a deeper insight can be obtained by studying correlations among these parameters. Figure~4 compiles such correlations in three panels: Fig.~4a plots $T$ versus $q$ for all centralities and particle species, Fig.~4b plots $T$ versus $\langle p_T\rangle$, and Fig.~4c plots $q$ versus $\langle p_T\rangle$. In Fig.~4a, we observe an inverse correlation between $T$ and $q$. Data points with higher $T$ tend to have lower $q$, and vice versa. This inverse relationship is most clearly driven by the centrality evolution discussed earlier. The trend in Fig.~~4a quantitatively confirms that the hottest systems (large $T$) are also the ones closest to equilibrium (small $q-1$), whereas cooler systems deviate more from equilibrium (larger $q-1$). In fact, $q$ can be related to an effective “entropy index” $n$ via $n = 1/(q-1)$ for Boltzmann-type distributions. A positive correlation between $T$ and $n$ (hence an inverse correlation between $T$ and $q$) was reported by Barnby \emph{et~~al.} in a study using Tsallis-like functions on STAR data \cite{barnby2009systematic}. Our results in Fig.~4a are in agreement with those findings and with other recent works that applied Tsallis fits in various systems \cite{yang2021dependence, badshah2023systematic, badshah2023evolution}. The implication is that higher excitation (higher $T$) coincides with more complete thermalization (lower $q$), reinforcing the physical picture of rapid equilibration in central heavy-ion collisions. Points corresponding to different hadron species at the same centrality also lie along this general $T$–$q$ trend, although there is some scatter due to the mass dependence; notably, for a given centrality, heavier particles (with higher $T$) indeed show slightly smaller $q$ values than lighter ones, as discussed.

\begin{figure}[t!]
\begin{center}
\hskip-0.153cm
\includegraphics[width=0.50\textwidth]{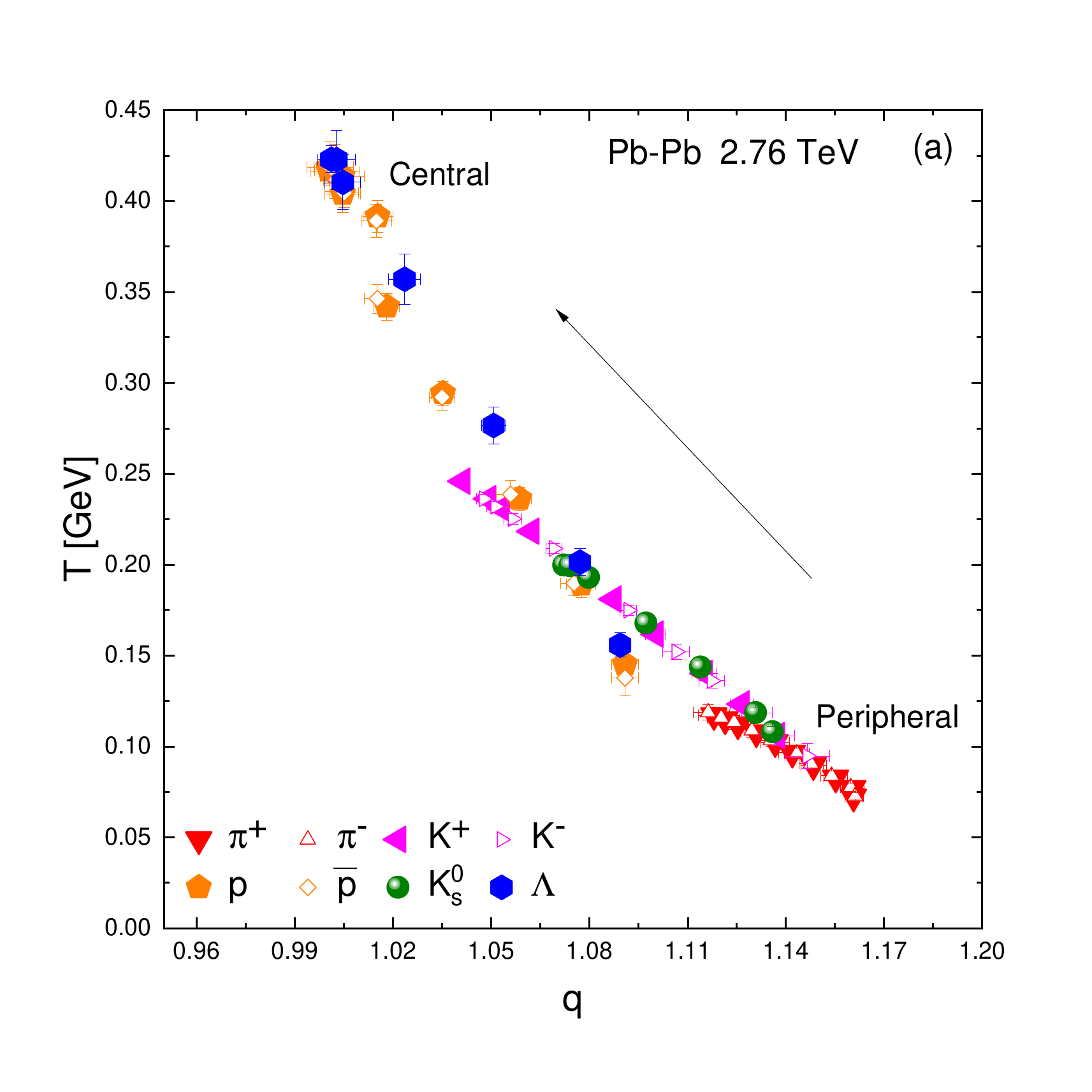}
\includegraphics[width=0.50\textwidth]{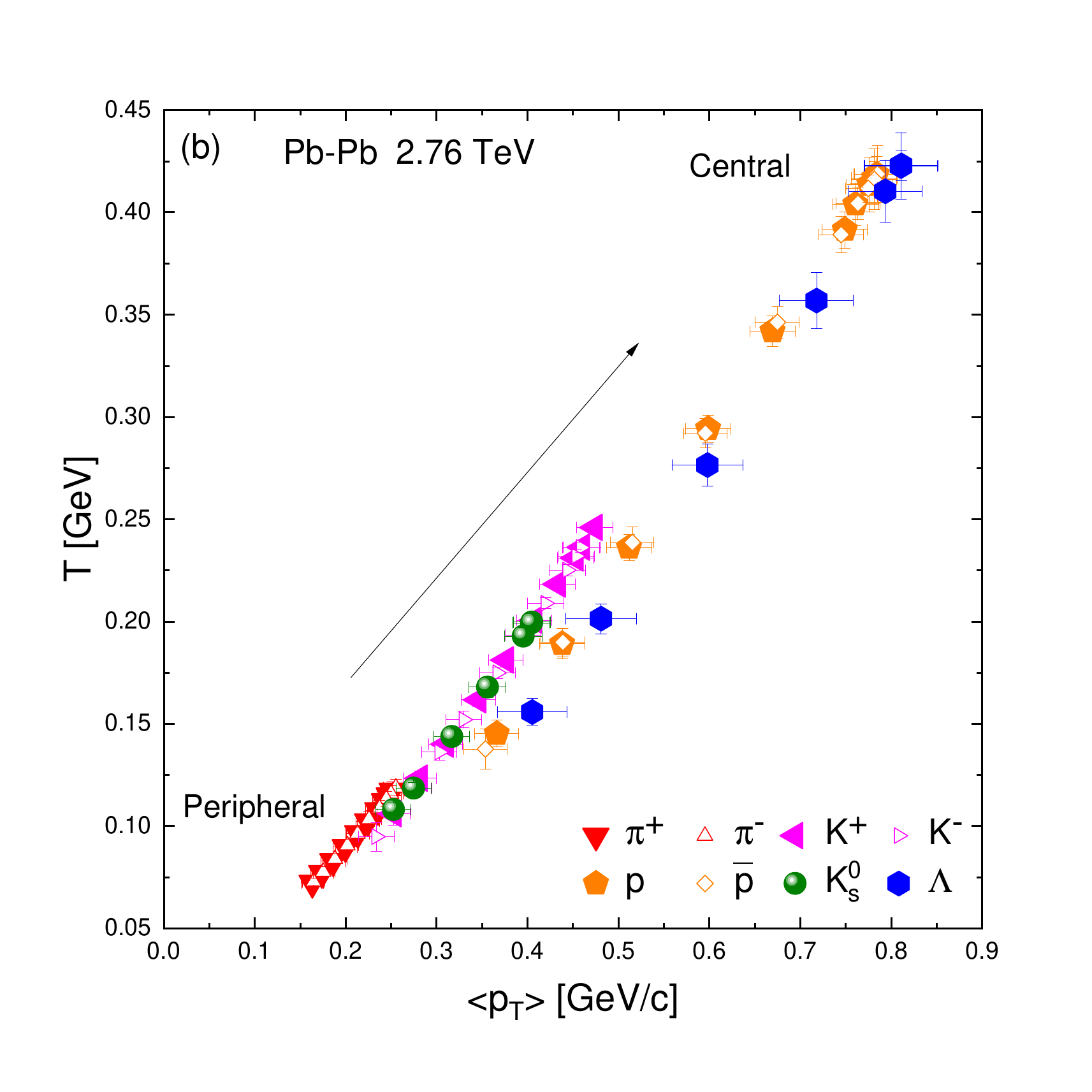}
\includegraphics[width=0.50\textwidth]{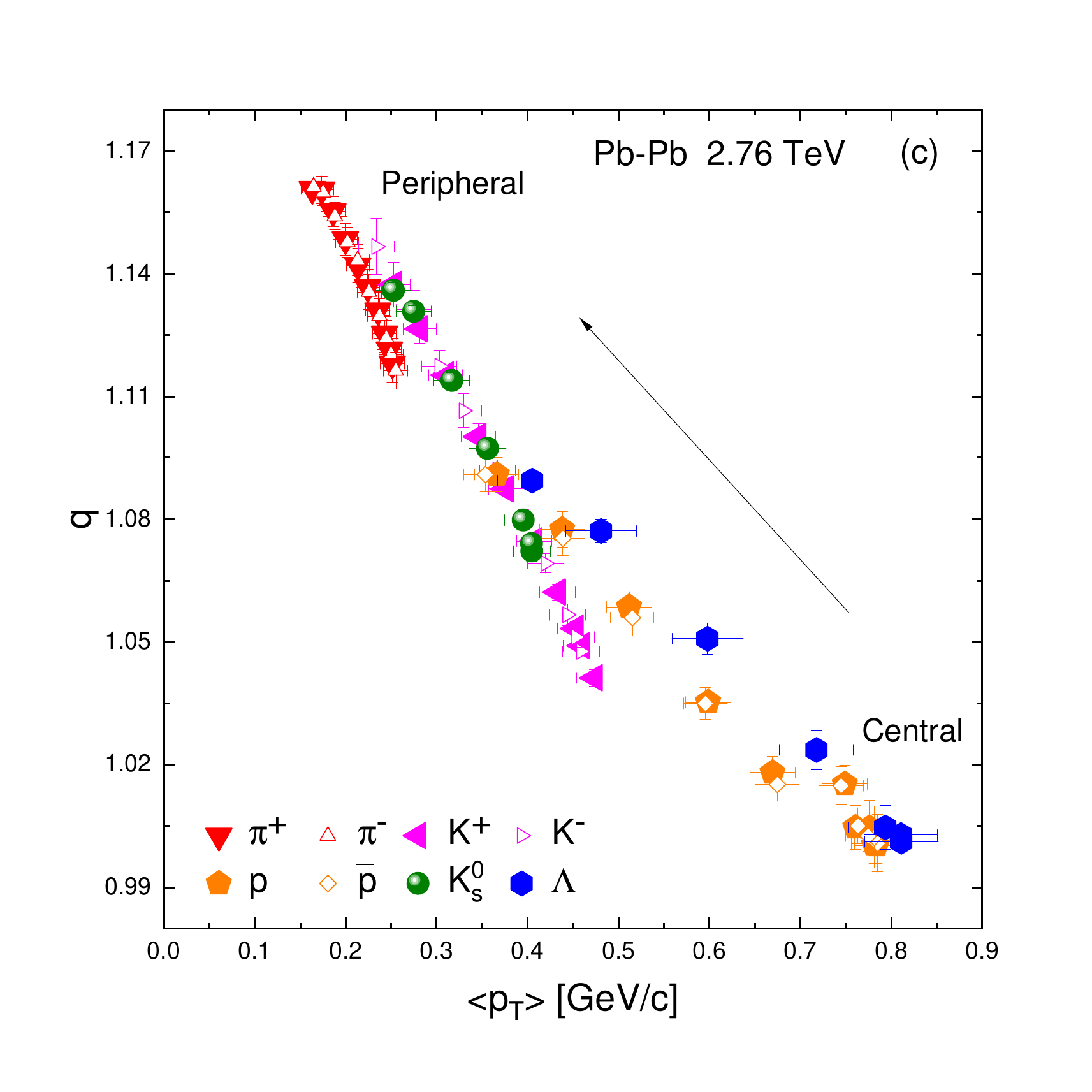}
\end{center}
\caption{Correlation between (a) $T$ and $q$, (b) $T$ and $\langle p_T\rangle$ and (c) $q$ and $\langle p_T\rangle$.}
\end{figure}

Figure~4b reveals a tight direct correlation between $T$ and $\langle p_T\rangle$. Since $\langle p_T\rangle$ is an integrated measure of the spectrum and $T$ controls the spectral slope, it is natural that events or particle species with a larger $T$ also have a larger mean $p_T$. The approximately linear correlation suggests that the Tsallis $T$ is a good proxy for the average momentum of particles: more energetic hotter systems push the mean of the $p_T$ distribution to higher values. In our dataset, each particle species separately shows an internal $T$–$\langle p_T\rangle$ correlation across centralities, and when all species are considered together, the correlation persists, with groups of points segregated by mass (heavier particles naturally have both higher $T$ and $\langle p_T\rangle$). This is conceptually consistent with hydrodynamic expectations and with empirical systematics observed at RHIC and LHC energies, where $\langle p_T\rangle$ rises with charged-particle multiplicity a proxy for centrality and tends to correlate with the spectra’s slope parameters.

Figure~4c plots $q$ against $\langle p_T\rangle$. We find an inverse correlation here as well: events with higher $\langle p_T\rangle$ which are the central, flowing systems have $q$ closer to 1, whereas those with lower $\langle p_T\rangle$ peripheral events exhibit larger $q$. Essentially, Fig.~4c recapitulates the message of Figs.~4a and 4b combined: since $T \propto \langle p_T\rangle$ and $T \propto 1/(q-1)$ approximately in our data. It follows that $\langle p_T\rangle$ should correlate inversely with $q$. This is indeed borne out: one can roughly fit an empirical relation of the form $T \propto \langle p_T\rangle \propto 1/(q-1)$ for our results. The correlation in Fig.~4c is essentially a consistency check, reinforcing that no contradictory trends are present among the three key parameters: higher mean momenta imply a closer-to-equilibrium state (lower $q-1$). This behavior is again indicative that collective flow (which raises $\langle p_T\rangle$) and strong rescattering (which lowers $q$) go hand-in-hand in central collisions. All these correlations highlight that the Tsallis parameters extracted from the spectra carry physically meaningful information about the system. Which implies that $T$ and $\langle p_T\rangle$ reflect the overall energy/momentum “push” in the event, while $q-1$ encapsulates deviations from thermal equilibrium. Their inter-relations are a manifestation of how these physical aspects are intertwined in heavy-ion collision dynamics.

\begin{figure}[t!]
\begin{center}
\hskip-0.153cm
\includegraphics[width=0.50\textwidth]{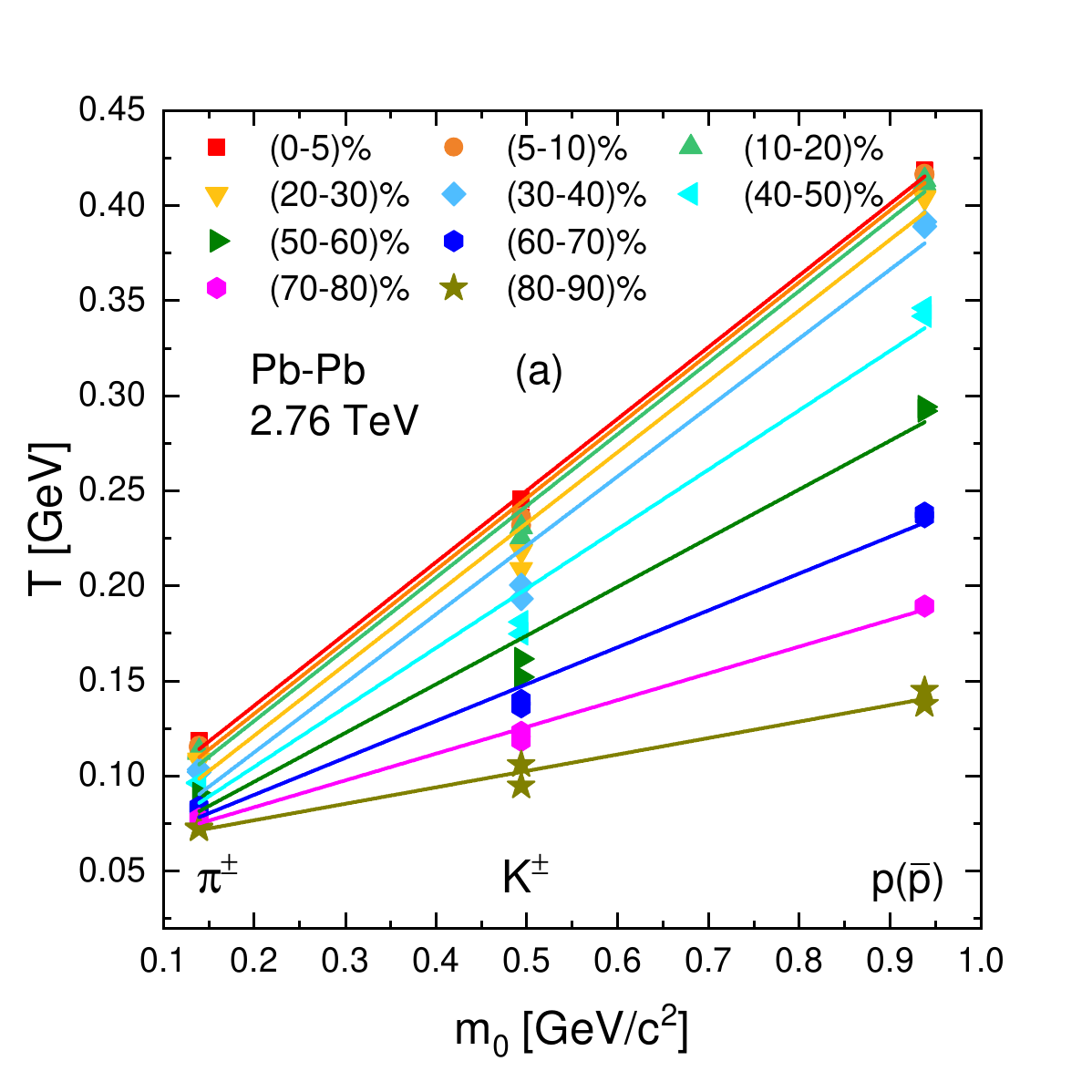}
\includegraphics[width=0.50\textwidth]{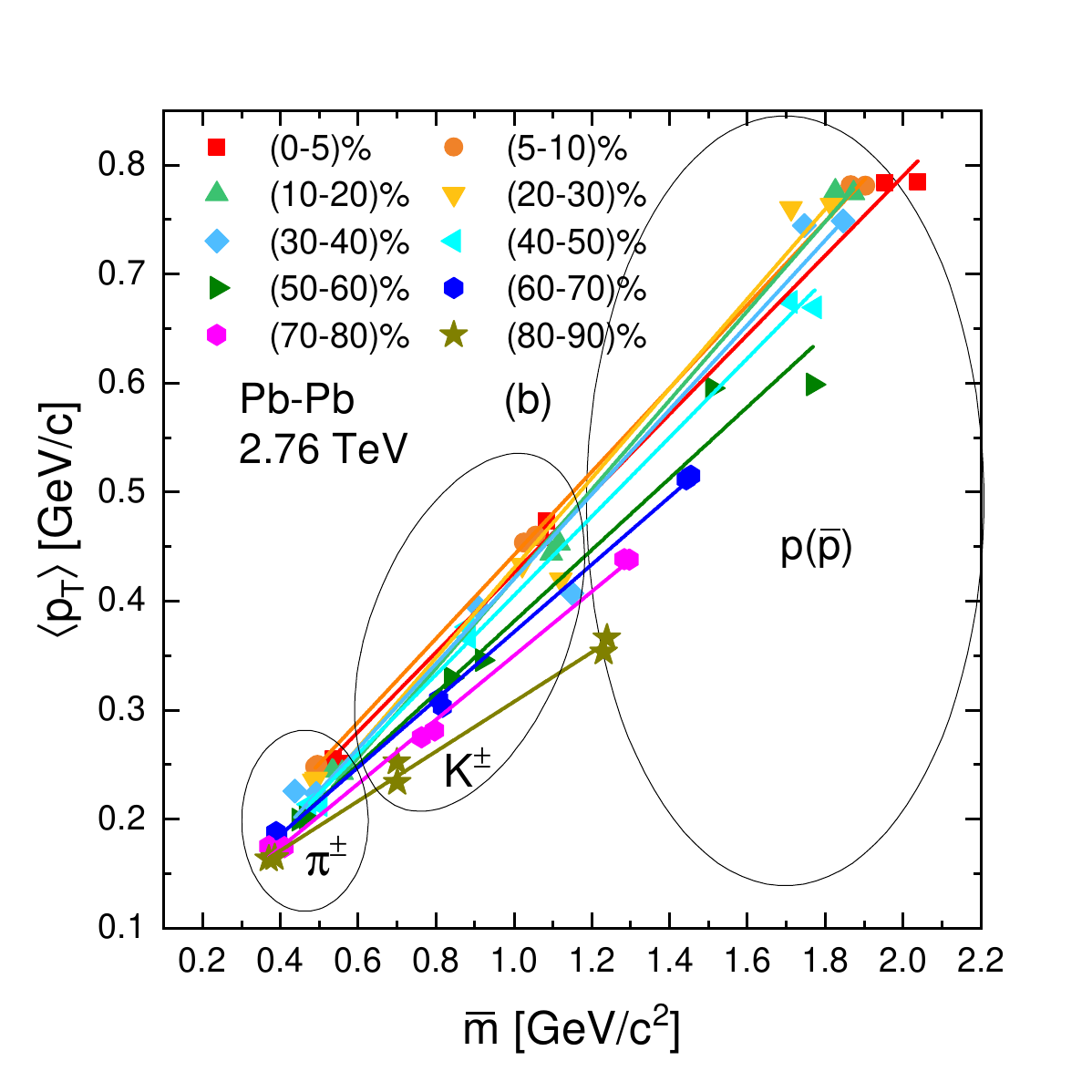}
\end{center}
\caption{(a) Effective temperature $T$ as a function of rest mass $m_0$ and (b) $\langle p_T\rangle$ versus mean moving mass $\overline{m}$ for light flavoured particles produced in Pb-Pb collisions at $\sqrt{s_{NN}}$ = 2.76 TeV. The solid lines are the linear fitting results.}
\end{figure}

Given that the Tsallis temperature $T$ includes contributions from both the random thermal motion of particles and collective transverse flow, it is desirable to disentangle these two effects. A higher mass particle having a larger $T$, as we observed, suggests that radial flow significantly influences the spectral shape. To isolate the pure thermal component of the motion, one can employ a procedure to extract the so-called kinetic freeze-out temperature $T_0$ the temperature of the system at the point of last inelastic collision and the average transverse flow velocity $\langle \beta_T \rangle$ from the systematic variation of $T$ and $\langle p_T\rangle$ with particle rest mass ($m_0$) and mean moving mass ($\overline{m}$), respectively. We followed an established approach \cite{wei2016disentangling, wei2016kinetic, badshah2024centrality} to perform this extraction, focusing on the pion, kaon, and proton data where the centrality intervals are identical. Figure~5 illustrates the procedure for a representative centrality class of Pb–Pb collisions at $\sqrt{s_{NN}}=2.76$~TeV. In Fig.~5a, the effective temperature $T$ obtained from Tsallis fits is plotted as a function of the rest mass $m_0$ of the particle (for $\pi$, $K$, $p$ in that centrality). We observe an approximately linear increase of $T$ with $m_0$, reflecting the fact that heavier particles exhibit higher slope parameters. We then perform a linear least-squares fit of $T$ vs $m_0$ relation using Eq. (\ref{eq:4a}). The intercept of this linear fit at $m_0=0$, i.e. extrapolating to a hypothetically zero-mass particle yields an estimate of $T_0$, the mean kinetic freeze-out temperature of the system \cite{wei2016kinetic, badshah2024centrality}. Physically, $T_0$ represents the temperature of the system in the absence of flow effects – it can be thought of as the true thermal temperature at freeze-out that lighter particles would have if they did not receive any flow boost. Meanwhile, Fig.~5b shows the mean transverse momentum $\langle p_T\rangle$ as a function of the particle’s mean moving mass $\overline{m}$ (often taken as $m_T$ averaged over the momentum distribution) for the same set of particles. This also exhibits an increasing linear trend, since heavier particles have higher $\langle p_T\rangle$. Fitting $\langle p_T\rangle(\overline{m})$ with a linear function given in Eq. (\ref{eq:4b}) allows us to extract the slope $\langle \beta_T \rangle$, which is interpreted as the average transverse flow velocity of the system \cite{wei2016disentangling}. In essence, $\langle \beta_T \rangle$ quantifies the common collective velocity kick imparted to particles of different masses. In our analysis, we find that the slope $a$ from the $T$–$m_0$ plot and the slope from the $\langle p_T\rangle$–$\overline{m}$ plot are numerically very similar for each centrality see Tables \ref{tab:2} and \ref{tab:3}, implying that either method yields a comparable $\langle \beta_T \rangle$. 

\begin{table}[!ht]
\caption{The values of y-intercept and slope extracted from the linear fit of the $T$ versus $m_0$ relationship through Eq. (\ref{eq:4a}).} \label{tab:2}
    \centering
    \begin{tabular}{cccccc}
     \hline
    \hline
        \textbf{Collision} & \textbf{Particles } & \textbf{Centrality} & \textbf{Intercept} & \textbf{Slope} & \textbf{$\chi^2$} \\ 
        \textbf{Energy} & \textbf{Types} & \textbf{Bins} & \textbf{($\langle T_0\rangle$ [GeV])} &   &  \\ 
        \hline
        \textbf{} & ~ & 0-5 \% & 0.0617$\pm$0.0031 & 0.3771$\pm$0.0089 & 0.0070 \\ 
        \textbf{} & ~ & 5-10 \% & 0.0617$\pm$0.0045 & 0.3771$\pm$0.0149 & 0.0119 \\ 
        \textbf{} & ~ & 10-20 \% & 0.0617$\pm$0.0040 & 0.3771$\pm$0.0129 & 0.0103 \\ 
        \textbf{} & ~ & 20-30 \% & 0.0617$\pm$0.0025 & 0.3771$\pm$0.0089 & 0.1100 \\ 
        \textbf{Pb---Pb} & $\pi^{\pm}$, $K^{\pm}$, $p(\bar{p})$ & 30-40 \% & 0.0501$\pm$0.0081 & 0.3608$\pm$0.0279 & 0.0223 \\ 
        \textbf{2.76 TeV} & ~ & 40-50 \% & 0.0455$\pm$0.0070 & 0.3126$\pm$0.2070 & 0.0166 \\ 
        \textbf{} & ~ & 50-60 \% & 0.0436$\pm$0.0060 & 0.2562$\pm$0.0175 & 0.0140 \\ 
        \textbf{} & ~ & 60-70 \% & 0.0404$\pm$0.0031 & 0.1940$\pm$0.0101 & 0.0081 \\ 
        \textbf{} & ~ & 70-80 \% & 0.0372$\pm$0.0016 & 0.1408$\pm$0.0050 & 0.0040 \\ 
        \textbf{} & ~ & 80-90\% & 0.0352$\pm$0.0071 & 0.0745$\pm$0.0173 & 0.0138 \\ 
        \hline
    \end{tabular}
\end{table}

\begin{table}[!ht]
\caption{The values of y-intercept and slope extracted from the linear fit of the $\langle p_T\rangle$ versus $\overline{m}$ relationship through Eq. (\ref{eq:4b}).} \label{tab:3}
    \centering
    \begin{tabular}{cccccc}
     \hline
    \hline
        \textbf{Collision} & \textbf{Particles } & \textbf{Centrality} & \textbf{Intercept} & \textbf{Slope} & \textbf{$\chi^2$} \\ 
        \textbf{Energy} & \textbf{Types} & \textbf{Bins} &  &  ($\langle \beta_T\rangle$ [c]) &  \\ 
        \hline
        \textbf{} & ~ & 0-5 \% & 0.0617$\pm$0.0054 & 0.3771$\pm$0.0087& 0.0010 \\ 
        \textbf{} & ~ & 5-10 \% & 0.0573$\pm$0.0068 & 0.3782$\pm$0.0109 & 0.0001 \\ 
        \textbf{} & ~ & 10-20 \% & 0.0537$\pm$0.0082 & 0.3770$\pm$0.0132& 0.0010 \\ 
        \textbf{} & ~ & 20-30 \% & 0.0468$\pm$0.0119 & 0.3727$\pm$0.0193 & 0.0056 \\ 
        \textbf{Pb---Pb} & $\pi^{\pm}$, $K^{\pm}$, $p(\bar{p})$ & 30-40 \% & 0.0400$\pm$0.0150 & 0.3628$\pm$0.0244 & 0.0070 \\ 
        \textbf{2.76 TeV} & ~ & 40-50 \% & 0.0425$\pm$0.0128 & 0.3126$\pm$0.0207 & 0.0008 \\ 
        \textbf{} & ~ & 50-60 \% & 0.0459$\pm$0.0108 & 0.2562$\pm$0.0174 & 0.0035 \\ 
        \textbf{} & ~ & 60-70 \% & 0.0514$\pm$0.0062 & 0.1940$\pm$0.0101 & 0.0002 \\ 
        \textbf{} & ~ & 70-80 \% &0.0555$\pm$0.0031 & 0.1408$\pm$0.0050 & 0.0002\\ 
        \textbf{} & ~ & 80-90\% & 0.0595$\pm$0.0039 & 0.0865$\pm$0.0063 & 0.0003 \\ 
        \hline
    \end{tabular}
\end{table}

This near-equality ($a \approx \langle \beta_T \rangle$) suggests that one could even use the mass-dependence of $T$ alone to estimate the flow velocity – a point noted in our results and also hinted at in previous work \cite{wei2016disentangling}. Nevertheless, using both $T(m_0)$ and $\langle p_T\rangle(\overline{m})$ relations in tandem cross-checks the consistency of the extracted $T_0$ and $\langle \beta_T \rangle$. We emphasize that in extracting $T_0$ and $\langle \beta_T \rangle$, we included only $\pi^\pm$, $K^\pm$, and $p(\bar{p})$ where the centrality selection was identical; the strange hadrons $K^0_S$ and $\Lambda$ were excluded from this particular step because their available centrality intervals e.g. 20–40\% do not exactly match the others. This ensures that our linear fits are not biased by mixing different centrality conditions. (Excluding $K^0_S$ and $\Lambda$ does not introduce any systematic uncertainty in the extraction, but simply means we focus on the light hadrons for consistency.)

Having obtained $T_0$ and $\langle \beta_T \rangle$ for each centrality class via the above procedure the numerical values are compiled in Tables \ref{tab:2} and \ref{tab:3}, we now examine how these freeze-out parameters vary with centrality. Figure~~6a shows the centrality dependence of $\langle T_0 \rangle$, the average kinetic freeze-out temperature. We find that $\langle T_0 \rangle$ increases from peripheral collisions up to mid-central collisions, and then saturates or plateaus for the most central collisions. Specifically, going from the most peripheral bin e.g. 80–90\% peripheral into central bins, $\langle T_0 \rangle$ rises sharply, reaching a maximum around the 20–30\% centrality class. For centralities finer than about 20–30\% toward 0\% central, i.e. head-on collisions, $\langle T_0 \rangle$ no longer grows and instead flattens out. In our data, this saturation value of $\langle T_0 \rangle$ is on the order of $\sim100$–110~~MeV exact values in Table \ref{tab:2}, after having risen from a peripheral value around 80–90~~MeV. The initial rise of $\langle T_0 \rangle$ with centrality can be understood as follows: more central collisions deposit greater energy density, which initially leads to a hotter kinetic freeze-out temperature (the system can cool to a lower temperature only in the very central collisions after substantial expansion). However, the observed saturation at mid-central to central events is a striking feature. One possible interpretation is that once the collisions become energetic enough beyond a threshold centrality, any additional energy pumped into the system goes into driving a phase transition or into expansion work rather than raising the kinetic freeze-out temperature. In other words, the system may be hitting the limit of hadronic temperature, beyond which extra energy is absorbed as latent heat for parton-hadron phase transition. The centrality interval of 20–30\% appears to correspond to the point where this change occurs in our analysis. It is tempting to associate this behavior with the onset of deconfinement or a softest-point effect in the equation of state. As the collision becomes more central, the fireball spends more time in a mixed phase QGP + hadrons or undergoes a soft phase transition such that the temperature ceases to increase despite higher input energy. While our experiment is at a fixed collision energy 2.76 TeV, a similar saturation of kinetic freeze-out $T_0$ across centrality was also hinted at in intermediate-energy collisions in Ref.~\cite{badshah2024centrality}, which the authors interpreted as a signature of approaching the QGP phase transition. Our observation here at the LHC suggests that even at the highest energies, central collisions might achieve a state where the kinetic freeze-out temperature is limited by phase transition dynamics. Any further increase in centrality simply results in more latent heat rather than exciting the system.

\begin{figure}[t!]
\begin{center}
\hskip-0.153cm
\includegraphics[width=0.50\textwidth]{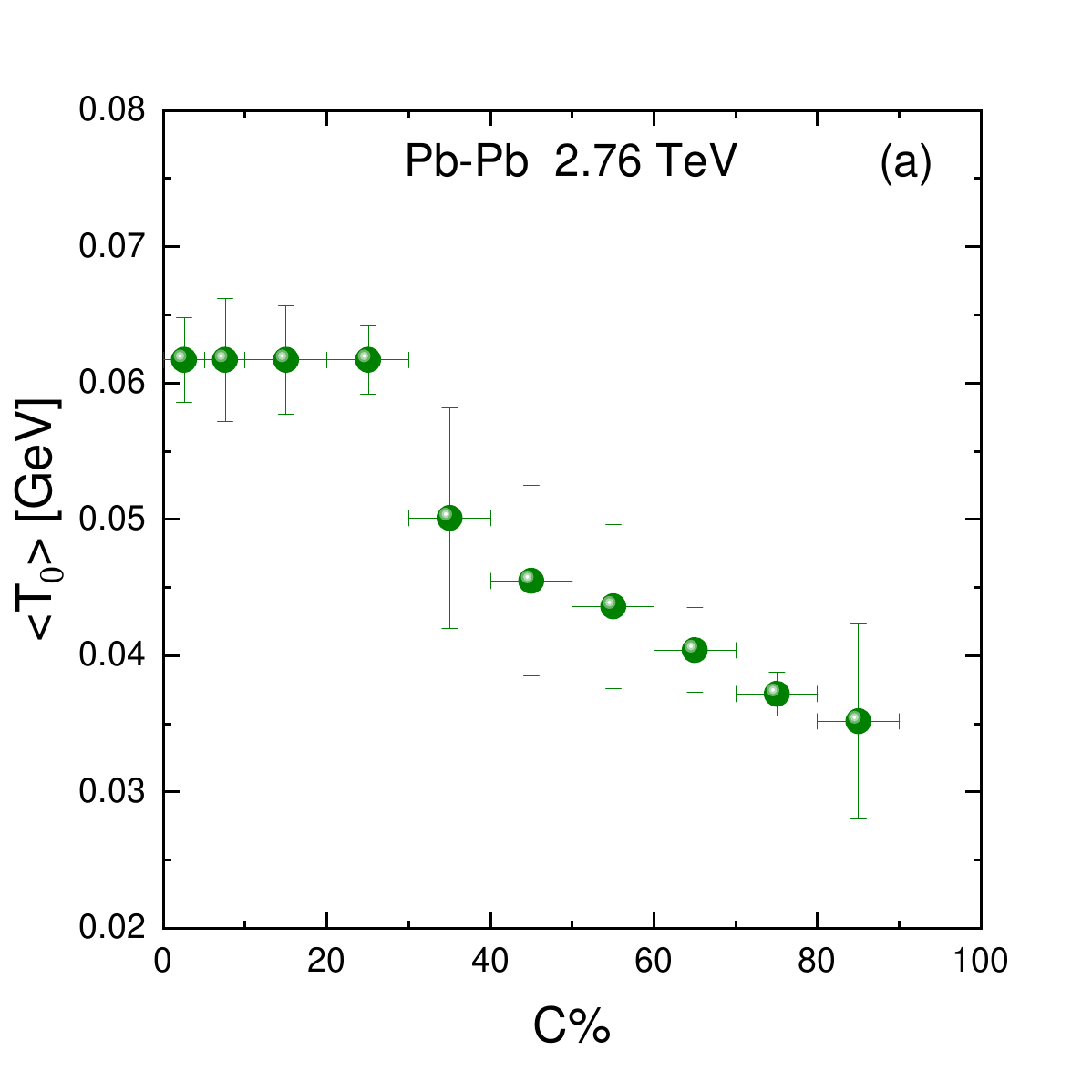}
\includegraphics[width=0.50\textwidth]{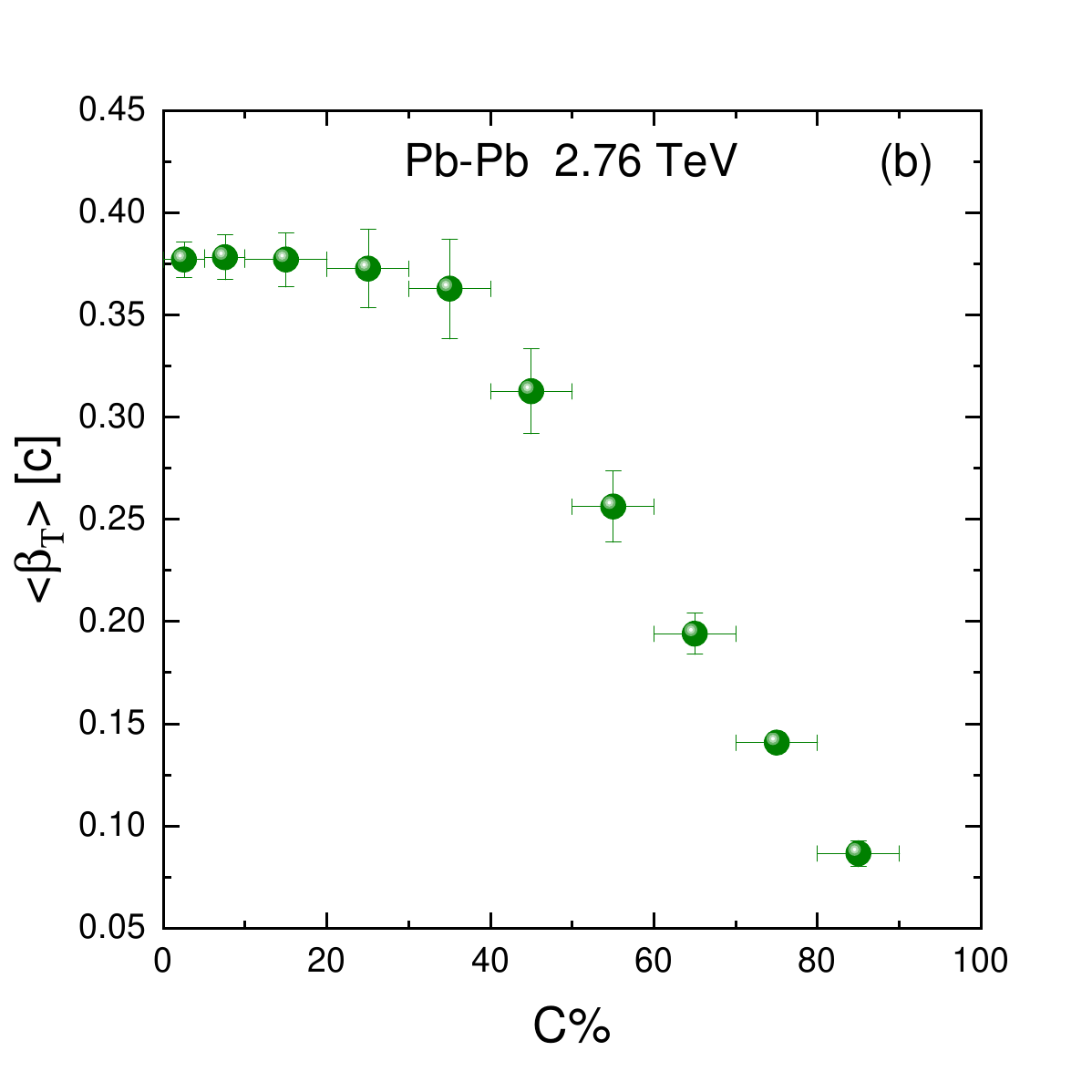}
\includegraphics[width=0.50\textwidth]{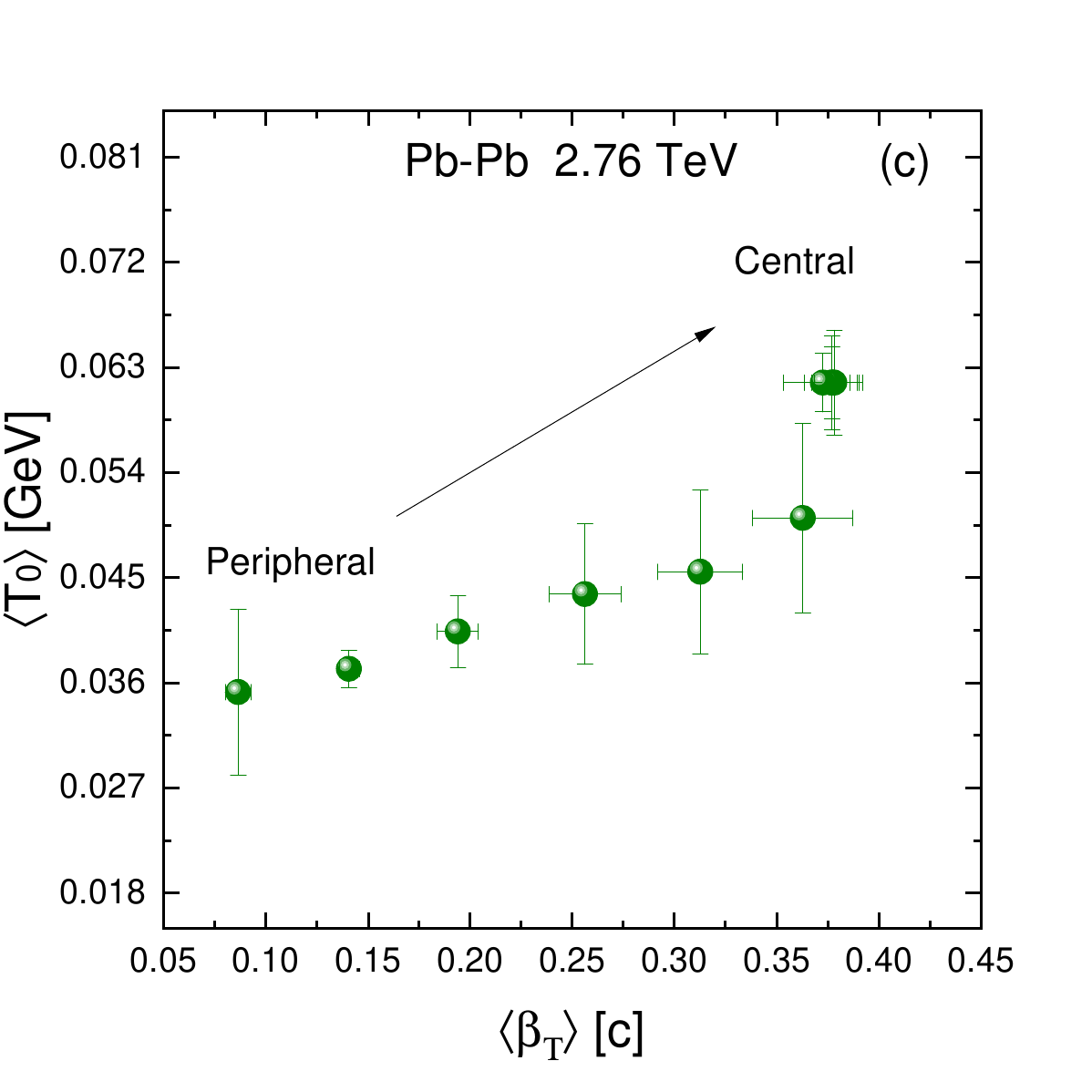}
\end{center}
\caption{(a) $\langle T_0 \rangle$ and (b) $\langle \beta_T \rangle$ as a function of centrality. (c) The correlation between $\langle \beta_T \rangle$ and $\langle T_0 \rangle$.}
\end{figure}

Figure~~6b shows the average transverse flow velocity $\langle \beta_T \rangle$ as a function of centrality. The $\langle \beta_T \rangle$ exhibits a very similar centrality trend to that of $T_0$. From peripheral to mid-central collisions, $\langle \beta_T \rangle$ rises rapidly, indicating that more central collisions generate significantly stronger collective flow. Quantitatively, $\langle \beta_T \rangle$ increases from around 0.3–0.35 in peripheral events to about 0.55–0.60 in semi-central events 20–30\% centrality in our fits again, see Table \ref{tab:3} for exact values. Beyond this centrality, moving to the top 10\% or 5\% most central, the flow velocity seems to saturate, plateauing at approximately $\langle \beta_T \rangle \sim0.6$. This means that the largest fireballs created in Pb–Pb collisions have an expansion speed that does not further increase with centrality once a certain size/particle-multiplicity is reached. The parallel between the $\langle \beta_T \rangle$ plateau and the $\langle T_0 \rangle$ plateau is notable. Both suggest that the 20–30\% centrality range is a turning point. A plausible explanation is that at 20-30\% centrality most of the energy is used as latent heat, leaving behind no fraction of energy to boost the system's transverse flow. The result is that beyond a certain centrality, the fireball’s temperature and collective velocity remain steady even as centrality increases, consistent with our observations. Our results for $\langle \beta_T \rangle$ in central collisions around 0.6 or 60\% of the speed of light are in line with values obtained from Blast-Wave fits at LHC energies \cite{source1, acharya2020production}, though the trend with centrality differs we will discuss this shortly. Finally, Fig.~6c directly correlates $\langle T_0 \rangle$ with $\langle \beta_T \rangle$ by plotting one versus the other for the set of centralities. We find a positive correlation: events with a higher average flow velocity also have a higher kinetic freeze-out temperature. This behaviour is consistent with pseudorapidity-differential Tsallis fits in $pp$ collisions, where $T_0$ and $\langle \beta_T \rangle$ vary coherently and are largest near mid-$\eta$ \cite{haifa2}, and with our multiplicity-differential analyses in symmetric/asymmetric nuclear and hadron–hadron collisions, which also show that $T_0$ and $\beta_T$ increase together with event activity \cite{Aj, Ajaz2023Symmetry}. In contrast, our blast-wave study of rapidity-dependent freeze-out at SPS energies finds that $T_{\text{kin}}$ changes with rapidity and particle mass while $\beta_T$ remains nearly rapidity independent and decreases with increasing mass, implying an effective anti-correlation between these parameters across hadron species \cite{Waqas2021Entropy}. In our present Tsallis-based extraction, however, both $T_0$ and $\beta_T$ rise together from peripheral to mid-central collisions, reflecting the simultaneous strengthening of flow and increase in effective freeze-out temperature up to the point of saturation. The roughly linear $\langle \beta_T \rangle$ vs.\ $T_0$ correlation (Fig.~6c) underscores that the hottest fireballs also expand the fastest, a sign of highly explosive events in central Pb–Pb collisions.
In our Tsallis-based extraction, however, both $T_0$ and $\beta_T$ rise together from peripheral to mid-central collisions, reflecting the simultaneous strengthening of flow and increase in effective freeze-out temperature up to the point of saturation. The roughly linear $\langle \beta_T \rangle$ versus $T_0$ correlation (Fig.~6c) underscores that the hottest fireballs also expand the fastest, a sign of highly explosive events in central Pb--Pb collisions.

It is noteworthy that the saturation of freeze-out parameters can only tentatively be linked to the possible phase change from hadronic to QCD matter. For a more definitive and exclusive conclusion in this regard, we need results from different lattice QCD and hydrodynamic models.  

The observed saturation of $\langle T_0 \rangle$ and $\langle \beta_T \rangle$ does signify the transition to a new dynamic regime, in which the system will attain thermal and collective equilibrium. Yet this saturation is not necessarily manifested in the effective Tsallis temperature ($T$) or other quantities which are directly derived from the Tsallis distribution, since these are the apparent spectral slopes of individual particle species and not the actual macroscopic freeze-out conditions. $T$ represents composite influences of thermal motion, collective flow, and non-extensivity, each of which depends on mass and momentum. Although the underlying thermal and flow components are saturated, small variations in spectral shape, particle composition, and extent of non-extensivity may, however, alter fitted slopes and give a smooth or monotonic behavior rather than an apparent plateau. Thus, the fact that the Tsallis parameters do not saturate clearly does not imply the new regime; instead, it is indicative of the fact that they are effective quantities sensitive to microscopic spectral detail as opposed to the global dynamical freeze-out state.

\section{Conclusion}
\label{secconc}
In this work, we have systematically studied the transverse momentum distributions of $\pi^{\pm}$, $K^{\pm}$, $p$($\bar{p}$), $K_s^0$, and $\Lambda$ in Pb--Pb collisions at $\sqrt{s_{NN}}=2.76$~TeV across multiple centrality classes. By applying the thermodynamically consistent Tsallis statistical framework to the $p_T$ spectra, we extracted the effective temperature $T$, non-extensivity parameter $q$, and calculated the mean transverse momentum $\langle p_T \rangle$ for each particle species and centrality interval.

The analysis reveals a clear centrality dependence of all extracted parameters. The effective temperature $T$ rises from peripheral to central collisions, reflecting stronger energy deposition and increased collective effects in the denser fireballs of central events. At the same time, $q$ systematically decreases toward central collisions, indicating that the system becomes progressively closer to local thermal equilibrium as the system size and participant density grow. An inverse correlation between $T$ and $q$ was observed, which signals rapid equilibration and strong thermalization in the most central collisions.

Our results further confirm a pronounced mass hierarchy. Heavier hadrons such as protons and $\Lambda$ exhibit higher $T$ and $\langle p_T \rangle$ compared to lighter particles. This mass ordering, persistent across all centralities, supports a multi-freeze-out scenario in which heavier hadrons decouple earlier from a hotter medium, while lighter particles remain coupled to the system until later stages when further cooling has occurred.

We also employed a two-step linear fit procedure to disentangle the thermal and collective components of the spectra, extracting the kinetic freeze-out temperature $\langle T_0 \rangle$ and mean transverse flow velocity $\langle \beta_T \rangle$. Both parameters increase rapidly from peripheral to mid-central events, and then reach a plateau beyond about 20–30\% centrality. This saturation of both $\langle T_0 \rangle$ and $\langle \beta_T \rangle$ in the most central collisions could be indicative of the system reaching the softest point of the equation of state or the onset of deconfinement, as additional energy input goes into expansion or phase change rather than further increasing the temperature or flow. The direct correlation between $\langle T_0 \rangle$ and $\langle \beta_T \rangle$ observed here suggests that, up to the saturation point, hotter fireballs also expand more rapidly.

The Tsallis-based approach provides a unified description of the $p_T$ spectra and reveals non-trivial patterns in freeze-out properties across centrality and particle species. The evidence for multi-freeze-out, non-equilibrium effects, and the possible signature of a phase transition in the most central events underscores the rich dynamics of heavy-ion collisions at the LHC. Further comparison with hydrodynamic and other statistical models, as well as future data at higher precision and for additional particle species, will help clarify the detailed nature of these transitions and the path to thermalization in QCD matter.

\begin{itemize}
\item {\bf Funding:} The present research work was funded by Princess Nourah bint Abdulrahman University Researchers Supporting Project number (PNURSP2026R106), Princess Nourah bint Abdulrahman University, Riyadh, Saudi Arabia.
\item {\bf Conflict of interest/Competing interests:} The authors declare no competing interests.
\item {\bf Ethics approval:} The authors declare their adherence to ethical standards concerning the content of this paper.
\item {\bf Consent to participate:} NA
\item {\bf Consent for publication:} All authors agreed to publish the current version of the manuscript after a thorough read.
\item {\bf Availability of data and materials:} The data analyzed in this manuscript is included in the manuscript and is cited at appropriate places as well.
\item {\bf Code availability:} The code can be provided on request. 
\item {\bf Authors' contributions:}  All authors contributed equally to this manuscript.
\item {\bf Acknowledgement:} We would like to acknowledge the support prvided by the Princess Nourah bint Abdulrahman University Researchers Supporting Project number (PNURSP2026R106), Princess Nourah bint Abdulrahman University, Riyadh, Saudi Arabia.
\end{itemize}

\bibliographystyle{aip}
\bibliography{newreferences}

\end{document}